\documentclass[showpacs,amsmath,amssymb,twocolumn,prx,superscriptaddress,10pt,aps,nofootinbib,maxbibnames=5]{revtex4-2}
\pdfoutput=1

\usepackage[utf8]{inputenc} 
\usepackage{tikz}
\usetikzlibrary{quantikz2}
\usepackage{graphicx}
\usepackage{siunitx}
\usepackage{amsmath,amssymb,amsthm,mathrsfs,amsfonts,dsfont}
\usepackage{subfigure, epsfig}
\usepackage{braket}
\usepackage{bm}
\usepackage{enumitem}
\usepackage[dvipsnames]{xcolor}
\usepackage{color}
\usepackage{fancybox, graphicx}
\usepackage[skins]{tcolorbox}
\usepackage{multirow}
\usepackage{mathtools}
\usepackage{makecell}
\usepackage{nowidow}

\usepackage{url}
\usepackage{breakurl}
\makeatletter
\g@addto@macro{\UrlBreaks}{\do\/\do\-\do\_\do\.\do\@}
\makeatother

\usepackage[colorlinks]{hyperref}
\hypersetup{
	colorlinks=true,
	citecolor = {blue},
    linkcolor=blue,
    filecolor=magenta,      
    urlcolor=cyan,
    breaklinks=true
}
\usepackage{cleveref}

\newcommand{\T}{\mathcal{T}}

\newcommand{\thm}[1]{\hyperref[thm:#1]{Theorem~\ref*{thm:#1}}}
\newcommand{\defn}[1]{\hyperref[defn:#1]{Definition~\ref*{defn:#1}}}
\newcommand{\lem}[1]{\hyperref[lem:#1]{Lemma~\ref*{lem:#1}}}
\newcommand{\prop}[1]{\hyperref[prop:#1]{Proposition~\ref*{prop:#1}}}
\newcommand{\fig}[1]{\hyperref[Fig:#1]{Figure~\ref*{Fig:#1}}}
\newcommand{\tab}[1]{\hyperref[tab:#1]{Table~\ref*{tab:#1}}}
\renewcommand{\sec}[1]{\hyperref[Sec:#1]{Section~\ref*{Sec:#1}}}
\newcommand{\append}[1]{\hyperref[App:#1]{Appendix~\ref*{App:#1}}}
\newcommand{\cor}[1]{\hyperref[cor:#1]{Corollary~\ref*{cor:#1}}}
\newcommand{\obs}[1]{\hyperref[obs:#1]{Observation~\ref*{obs:#1}}}

\definecolor{amethyst}{rgb}{0.6, 0.4, 0.8}

\newcommand{\norm}[1]{\left\lVert#1\right\rVert}
\newcommand{\vertiii}[1]{{\left\vert\kern-0.25ex\left\vert\kern-0.25ex\left\vert #1
		\right\vert\kern-0.25ex\right\vert\kern-0.25ex\right\vert}}

\def\ketbra#1{ |{#1}\rangle\!\langle{#1}| }

\usepackage{mathrsfs}

\allowdisplaybreaks

\usepackage{titlesec}
\begin{document}

\title{The Fast for the Curious: How to accelerate fault-tolerant quantum applications}

\author{Sam McArdle}
\email{sam.mcardle.science@gmail.com}
\affiliation{AWS Center for Quantum Computing, Pasadena, CA 91125, USA}

\author{Alexander M. Dalzell}
\affiliation{AWS Center for Quantum Computing, Pasadena, CA 91125, USA}

\author{Aleksander Kubica}
\affiliation{AWS Center for Quantum Computing, Pasadena, CA 91125, USA}
\affiliation{Yale Quantum Institute \& Department of Applied Physics, Yale University, New Haven, USA}

\author{Fernando G.S.L. Brand\~{a}o}
\affiliation{AWS Center for Quantum Computing, Pasadena, CA 91125, USA}

\begin{abstract}
\noindent We evaluate strategies for reducing the run time of fault-tolerant quantum computations, targeting practical utility in scientific or industrial workflows. Delivering a technology with broad impact requires scaling devices, while also maintaining acceptable run times for computations. Optimizing logical clock speed may require moving beyond current strategies, and adopting methods that trade faster run time for increased qubit counts or engineering complexity. We discuss how the co-design of hardware, fault tolerance, and algorithmic subroutines can reduce run times. We illustrate a selection of these topics with resource estimates for simulating the Fermi-Hubbard model.
\end{abstract}

\maketitle

\section{Introduction}
Computational methods are used in many scientific and industrial workflows, where they must satisfy constraints on their accuracy and cost (time and/or budget). In many cases, there are a number of classical algorithms available, offering a tradeoff between accuracy and cost. High-accuracy classical methods---the natural point of comparison for quantum algorithms---often cannot satisfy the time constraints. Instead, faster, heuristic or approximate classical algorithms are commonly used. A prominent example is the widespread use of density functional theory in industrial chemistry and materials science~\cite{deglmann2015IndustryChem}.

When searching for quantum applications, we may envisage the following positive scenarios: 1) quantum algorithms are comparably accurate and asymptotically more efficient than the classical algorithm used in practice, or 2) quantum algorithms are asymptotically less efficient than the classical algorithm used in practice, but provide a better solution which justifies a higher cost.\footnote{We will not discuss a third positive scenario, where the quantum algorithm is efficient for a task with no known efficient classical algorithms, thus enabling entirely new workflows. Example algorithms currently lacking practical applications include algorithms for algebraic problems~\cite{childs2010QAlgosForAlgebraicProblems}, algorithms for specific topological properties in restricted parameter regimes/settings~\cite{laakkonen2025JonesPolynomial},\cite[Sec.9.4]{dalzell2023E2E}, or random circuit sampling~\cite{hangleiter2023RandomCircuitSamplingRev}.} The first case is desirable, as it means quantum algorithms can be swapped into existing workflows. However, the prevalence and efficiency of classical heuristics makes it challenging to achieve.\footnote{See Refs.~\cite{Babbush2023Dynamics,rubin2023FusionDynamics} for a promising counterexample where the exact quantum algorithm provides speedups over the classical approximate method.} The second scenario---an example of which is computing ground state energies of molecules using quantum phase estimation~\cite{lee2022isThereEvidenceChemistry}---is more plausible, but is harder to evaluate because it results in unpredictable disruption of the existing workflow. Moreover, the practicality of either scenario is determined not by asymptotic costs, but by the observed run time--accuracy tradeoffs in real, finite-sized problem instances.

\begin{table*}[]
    \centering
    \begin{tabular}{c|c|c|c|c}
       \textbf{Problem} & \textbf{Logical qubits} & \textbf{T/Toffoli gates} & \textbf{Physical qubits} & \textbf{Run time} \\ \hline \hline
        \makecell{Ground state energy of molecule~\cite{low2025fastSOSSA} \\ Ground state energy of material~\cite{rubin2023MaterialsSim}} & \makecell{$10^3$ \\ $10^5$--$10^6$} & \makecell{$10^{9}$ \\ $10^{12}$--$10^{14}$} & \makecell{$2.2 \times 10^6$ \\ $ 2.8 \times 10^9 $ } & \makecell{$8$ hours$^{(P)}$ \\ 14 years$^{(P)}$ } \\ \hline
        \makecell{Time evolve spin system~\cite{Beverland2022Requirements} \\ SU$(3)$ lattice gauge theory dynamics~\cite{rhodes2024ExponentialLatticeGauge} \\ Kinetic energy loss in nuclear fusion$^\dag$~\cite{rubin2023FusionDynamics}} & \makecell{$10^2$ \\ $10^7$ \\  $10^3$--$10^4$} & \makecell{$10^5$ \\ $10^{21}$ \\ $10^{13}$--$10^{17}$} & \makecell{ $8.7 \times 10^4 $ \\ $ 1.1 \times 10^{11} $ \\ $ 2.5 \times 10^7 $} & \makecell{2 seconds$^{(P,M)}$ \\ $2 \times 10^9$ years$^{(P,M)}$ \\ $1.3 \times 10^3 $ years } \\ \hline
        \makecell{Options pricing~\cite{chakrabarti2021threshold} \\ Portfolio optimization$^*$~\cite{dalzell2022socp}} & \makecell{$10^4$ \\ $10^6$} & \makecell{$10^{10}$ \\ $10^{29}$} & \makecell{ $2.9 \times 10^7$ \\ $ 1.7 \times 10^{10} $} & \makecell{ $3.7$ days$^{(P)}$ \\ $ 2.4 \times 10^{17} $ years$^{(P)}$ } \\ \hline
        \makecell{Radar cross section$^*$~\cite{scherer2017concrete} \\ Simple fluid dynamics~\cite{penuel2024feasibilityAcceleratingIncompressible}} & \makecell{$10^8$ \\ $10^3$--$10^5$} & \makecell{$10^{29}$ \\ $10^{20}$--$10^{24}$} & \makecell{$1.9 \times 10^{12}$ \\ $ 9.7 \times 10^{7} $} & \makecell{ $2.5 \times 10^{17}$ years$^{(P,M)}$ \\ $1.8 \times 10^{10}$ years$^{(P,M)}$} \\ \hline
        \makecell{Factoring for RSA-2048 (optimize time)$^\dag$~\cite{gidney2021HowToFactor} \\ Factoring for RSA-2048 (optimize space)$^\dag$~\cite{gidney2025factor} \\ Discrete logarithm for EC-256$^\dagger$~\cite{litinski2023EllipticCurvesBaseline}} & \makecell{$10^4$ \\ $1.4 \times 10^3$ \\ $10^3$} & \makecell{$3 \times 10^9$ \\ $7 \times 10^8$ (per circuit) \\ $4 \times 10^7$} & \makecell{$2.9 \times 10^7$ \\ $3.1 \times 10^6 $ \\ $1.9 \times 10^6 $} & \makecell{$1.1$ days \\ $ 2 $ days \\ $17 $ minutes} \\ \hline
    \end{tabular}
    \caption{A selection of resource estimates for problems in chemistry, physics simulation, finance, differential equation solving, and cryptography.\footnote{Calculated physical resources may differ from the original publication, which may use different assumptions and methods in their resource estimates. For example, Ref.~\cite{gidney2025factor} achieves an estimate of less than one million physical qubits for factoring 2048-bit RSA integers (more than $3\times$ better than what is reported in the table) by using less than $1.5 \times$ routing/distillation overhead and densely storing a large fraction of the logical qubits in  ``cold storage'' by concatenating the surface code with yoked codes \cite{gidney2025yoked}. } Resource estimates are based on a minimal footprint surface code architecture (physical error rate $p=10^{-3}$), such that one $T$/Toffoli gate is implemented per $d$ syndrome extraction rounds, where $d$ is the code distance~\cite{litinski2019gameofsurfacecodes}. Both the time for a round of surface code syndrome extraction, and the reaction time (see Sec.~\ref{Subsubsec:ReactionTime}) are assumed to be \SI{1}{\micro \second}.
    We use $^\dag$ to denote that state preparation and/or measurement overheads are included in the estimate. Otherwise, run times are quoted for a single repetition of the circuit, which would need to be re-run to measure non-commuting observables (denoted by $^{(M)}$) or sweep over different parameters (denoted by $^{(P)}$). The number of repetitions required for $(P), (M)$ depends on the application considered. For example, one may need to sweep over tens or hundreds of molecular geometries if using ground state energies to compute reaction pathways. Measurement of observables to precision $\epsilon$ requires $\Omega(1/\epsilon^2)$ incoherent repetitions. Alternatively, one could use amplitude estimation, which increases $T$/Toffoli count and run time by $\Omega(1/\epsilon)$.
    Physical qubits and circuit run time estimates assume the midpoint of quoted ranges for logical qubits and $T$/Toffoli gates. We do not report physical qubits used for magic state distillation, and assume a routing overhead of $1.5 \times$. $^*$ denotes that the resource estimate could be substantially improved using techniques developed since the original publication. These resource estimates are likely to further improve, and differing assumptions between estimates make them incomparable. 
    }
    \label{tab:ResourceEstimates}
\end{table*}

Building a fault-tolerant quantum computer (FTQC) would enable the development and testing of heuristic quantum algorithms. In the meantime, we are limited to predicting the run times of the subset of quantum algorithms with provable accuracy guarantees. We can compile these algorithms to the primitive instructions for a proposed FTQC architecture to estimate the number of physical qubits required, as well as the run time of the given quantum circuit~\cite{Beverland2022Requirements}. Current proposals for building FTQCs predict logical clock speeds in the kHz--MHz range, much slower than the GHz clock speeds of classical computers. To claw back this speed disadvantage, it will be necessary to tailor computations to the underlying hardware platform. For the sake of simplicity, many existing resource estimates forgo this optimization, and proceed with standard models of FTQC, such as surface code quantum error correction and 2D planar connectivity. Moreover, since the first FTQCs will have a limited number of qubits, current resource estimates prioritize maintaining low physical qubit counts. This is achieved by applying logical gates sequentially, which increases algorithmic run times. A number of representative resource estimates are shown in~\cref{tab:ResourceEstimates}. We observe that, with a few notable exceptions, many applications are impractically slow under these assumptions. Moreover, it is also important to account for the overheads of repeating a quantum circuit many times in order to solve a given end-to-end problem~\cite{aaronson2015ReadTheFinePrint,dalzell2023E2E}. We define the total time to solve the end-to-end problem as the wall time (classical pre- and post-processing, plus the number of repetitions multiplied by the circuit run time). In some applications, we must repeat quantum circuits in order to sweep over a range of parameters or initial conditions. Similar repetitions are required for classical algorithms. However, quantum circuits also require additional overhead to measure non-commuting observables, if required for the application~\cite{huang2020predicitingmanypropfewmeas,huggins2022ExpectationValue,apeldoorn2022TomographyStatePreparationUnitaries}. Such an overhead is not required in classical algorithms, where information can be cloned and stored. Once these overheads are accounted for, the most promising end-to-end applications may take days, even with optimistic assumptions. It is clear that slow logical clock speeds could limit the practicality of many quantum applications.

Faster logical clock speeds will be desirable beyond the need to compete with classical computers. There are many examples of winner-takes-all phenomena in classical computing, such as real-time weather forecasting for financial applications, or computational design of pharmaceuticals or engineering projects. In these cases, users are willing to pay a premium for faster results in time-sensitive calculations. FTQC resource estimates should not be one-size-fits-all, neither for applications nor hardware. Rather, we advocate for an approach that estimates the required logical clock speed by working backward from the workflow constraints. We expect that reaching the specified logical clock speed will require targeted optimizations at the algorithmic, compilation and FTQC levels, which must be tailored for the underlying qubit modality.

\section{Overview}

\begin{figure*}
    \centering
\includegraphics[width=0.95\linewidth]{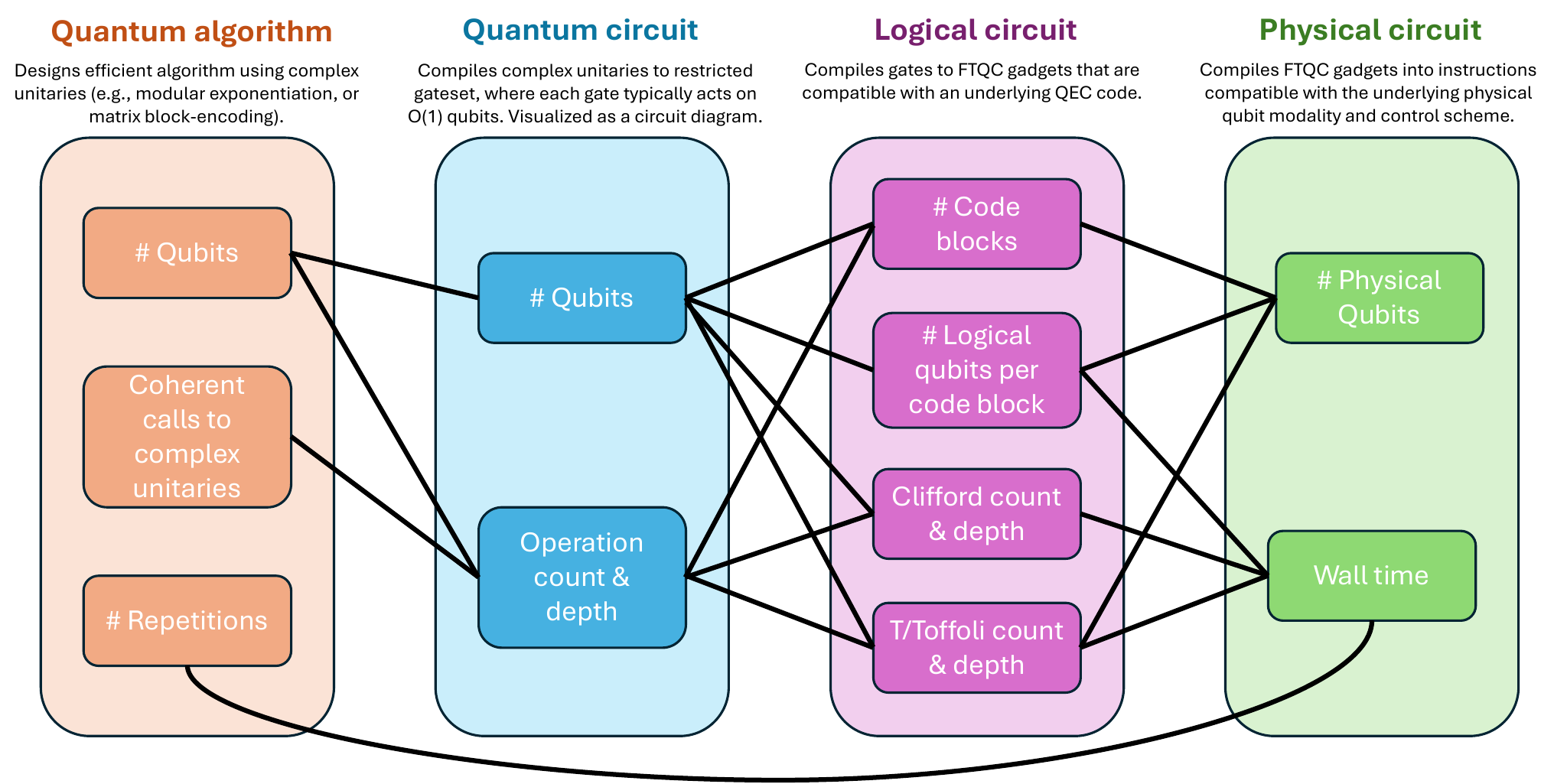}
    \caption{
An abstract illustration of the compilation stack from a quantum algorithm (e.g., Shor's algorithm) to its physical implementation. The stack is an artificial division of choices that has developed over time to help explore the space of all possible designs. Compilation can take place within layers (e.g., trading coherent circuit depth for algorithm repetitions) or between layers (e.g., trading qubits to reduce $T$-count in QROAM, see Sec.~\ref{Subsec:LookupTablesQROM}). The connections between layers signify that information must be passed between layers in order to optimize the compilation. For example, optimizing the $T$-depth in the logical layer depends on the spacetime cost of preparing magic states in the physical layer. The dependencies between layers means that compilation is an iterative, rather than one-way process. }
    \label{fig:AbstractCompilation}
\end{figure*}

Motivated by the requirements of practical quantum applications, this perspective aims to: 1) collect and organize several strategies for reducing the wall time of quantum computations, 2) discuss their benefits and drawbacks, and 3) quantitatively illustrate the interplay of the available tradeoffs in a concrete example. 

At a high-level, we seek to run a given quantum algorithm on a realizable FTQC architecture---with the overall goal of either minimizing the wall time subject to a limit on the cost, or vice versa. This provides a well-defined logical clock speed that depends on choices made in different layers of the computational stack. 
As shown in Fig.~\ref{fig:AbstractCompilation}, the stack layers are tightly coupled and are often in tension with each other, and so benefit from a co-design mindset. Over-optimizing one layer of the stack may lead to sub-optimal results. For example, the optimal quantum error-correcting (QEC) code from a storage perspective may be difficult to physically realize, or may have slow schemes for logical computation. As a result, it is typical to search for methods that compromise between the requirements of different layers of the stack. To facilitate engineering, we may also prioritize other `hardware friendly' features, such as compatibility with modular architecture designs, or simple control schemes.

There are a number of ways to measure and specify the logical clock speed of an FTQC. One approach is to calculate a peak theoretical logical clock speed by assuming parallel logical operations and multiplying the number of logical qubits by the speed of each logical operation. This methodology provides an application-agnostic value that depends only on hardware parameters and the choice of FTQC scheme. However, as we will discuss in this perspective, running quantum algorithms introduces a number of additional nuances which may limit the attainable logical clock speed, such as algorithmic \& compilation parallelism, as well as bottlenecks in magic state consumption, operation routing, \& communication. As such, we can define an effective logical clock speed as the average number of elementary logical quantum operations performed per second in a circuit compiled for the FTQC. This is analogous to the way floating-point operations per second (flop/s) are presented in classical computing, where it is common to specify both a peak theoretical performance, and a maximum practical performance extracted from benchmarking. We note that each logical quantum operation is not in one-to-one correspondence with a classical floating-point operation. Each floating-point operation is composed of a number of elementary logical operations (AND, OR, NOT) and many elementary logical operations can be performed all within a single CPU clock cycle (e.g., in classical arithmetic logic units). Furthermore, by performing floating-point operations in parallel across many processors, the largest classical supercomputers are capable of reaching the exascale regime ($10^{18}$ flop/s), and graphics processing units (GPUs) can achieve more than $10^{13}$ flop/s even on a single chip. Currently foreseeable FTQCs will be limited to significantly smaller effective logical clock speeds because: 1) each logical operation requires a number of physical operations (which in many qubit modalities are slower than operation of classical transistors), 2) each quantum logical operation is comparable to an elementary classical gate rather than to a floating-point operation, and 3) each logical operation requires simultaneous classical processing to control the system and perform quantum error correction.\footnote{As a result, even if physical quantum operations could be performed at rates comparable to classical clock speeds (e.g., physical operations in photonic platforms are characterized by GHz clock rates~\cite{bombin2021interleaving}), the quantum clock speed may need to be slowed down to avoid a decoding backlog~\cite{terhal2015QECforQuantumMemories}.} We refer the reader to Ref.~\cite{babbush2021FocusBeyondQuadratic} for a detailed discussion of these overheads.  A visual comparison of the resulting gap in estimated timescales between classical and quantum operations is depicted in Fig.~\ref{Fig:ClassicalVsQuantumTimescales}.

\begin{figure*}[t]
\centering
\begin{tikzpicture}

\definecolor{deepurple}{RGB}{88, 24, 69}
\definecolor{goldyellow}{RGB}{255, 190, 11}
\definecolor{mutedpurple}{RGB}{146, 111, 159}
\definecolor{warmyellow}{RGB}{255, 200, 87}
\definecolor{brightpurple}{RGB}{128, 0, 128}  
\definecolor{brightyellow}{RGB}{255, 223, 0}  

\def\textSize{\normalsize}
\def\labelSize{\footnotesize}

\def\tikzVspace{0.56cm}        
\def\tikzVlinespace{1.0cm}    
\def\tikzTextSep{0.1cm}     
\def\tikzTextcolwidth{7.8cm}  
\def\tikzNumvlines{10}       
\def\tikzTopspace{0.2cm}    
\def\tikzHoffset{0.3cm}     
\def\tikzHoffsetRight{0.4cm}     
\def\tikzCircleLabelOffset{0.1cm}     
\def\tikzCircleRadius{0.15cm}   
\def\numRows{5}   

\def\tikzHlinewidth{\tikzHoffset+\tikzHoffsetRight+\tikzNumvlines*\tikzVlinespace-\tikzVlinespace}    

\def\xShift{0 cm}

\def\vlabels{%
    {},
    {\SI{1}{\pico \second}},%
    {},
    {},
    {\SI{1}{\nano \second}},%
    {},
    {},
    {\SI{1}{\micro \second}},
    {},
    {}
}

\foreach [count=\i starting from 1] \label in \vlabels {
    \pgfmathsetmacro{\x}{(\i-1)}
    \draw[dotted] (\xShift+\tikzHoffset+\x*\tikzVlinespace,\tikzTopspace) -- 
        (\xShift+\tikzHoffset+\x*\tikzVlinespace,{-(\numRows-1)*\tikzVspace-\tikzTopspace});
    \node[anchor=south,rotate=0] at (\xShift+\tikzHoffset+\x*\tikzVlinespace,\tikzTopspace+0.1cm) {\textSize \label};
}

\foreach [count=\y starting from 0] \colorVal/\x/\len/\pow/\circlelabel in {%
warmyellow/{Average time per flop for 10 teraflop/s GPU}/1/-4/{\SI{100}{\femto \second}},%
warmyellow/{Time for single classical gate}/2.5/-3/{\SI{2.5}{\pico \second} \cite{stillmaker2017scalingEquationsCMOS}},%
    warmyellow/{Clock cycle time for 3 GHz classical processor}/3.3/-1/{\SI{330}{\pico \second}},%
    mutedpurple/{Time for two-qubit physical quantum gate}/4.2/1/{\SI{42}{\nano \second} \cite{google2025Below}},%
    mutedpurple/{Estimated time for  logical quantum gate}/2.8/4/{\SI{28}{\micro \second}}%
    } {
    \node[anchor=east, text width=\tikzTextcolwidth, text ragged left] 
        at (\xShift-\tikzTextSep,-\y*\tikzVspace) {\textSize \x};
    
    \draw[black] (\xShift-\tikzTextcolwidth,-\y*\tikzVspace + \tikzVspace/2) -- (\xShift+\tikzHlinewidth,-\y * \tikzVspace + \tikzVspace/2);
    
    \pgfmathsetmacro{\logvalueAdjusted}{(4+\pow+log10(\len))*\tikzVlinespace}
    \draw[fill=\colorVal] (\xShift+\tikzHoffset+\logvalueAdjusted,-\y*\tikzVspace) circle (\tikzCircleRadius);
    
    \node[anchor=west, font=\scriptsize] at (\xShift+\tikzHoffset+\logvalueAdjusted+\tikzCircleLabelOffset,-\y*\tikzVspace) {\labelSize \circlelabel};

}
\draw[black] (\xShift-\tikzTextcolwidth,-\numRows*\tikzVspace + \tikzVspace/2) -- (\xShift+\tikzHlinewidth,-\numRows * \tikzVspace + \tikzVspace/2);

\end{tikzpicture}



\caption{\label{Fig:ClassicalVsQuantumTimescales}A comparison between timescales in classical processors (yellow) and the anticipated timescales of quantum processors (purple), based on currently achievable benchmarks in superconducting platforms \cite{google2025Below} designed for compatibility with the 2D surface code. Classical gate time corresponds to the delay for a fan-out of 4 (FO4) CMOS inverter gate, which acts as a standard benchmark---the reported value is taken from simulations of the \SI{7}{\nano\metre} process \cite{stillmaker2017scalingEquationsCMOS}. Logical quantum gate time estimate assumes $d=25$ rounds of syndrome extraction (for lattice surgery), with each round taking \SI{1.1}{\micro \second}, consistent with the cycle time reported in Ref.~\cite{google2025Below}.}
\end{figure*}

A recurring theme in this perspective is the ability to trade qubits for time (`spacetime tradeoffs'). Within limits, spacetime tradeoffs can be used to reduce the run time (and potentially total spacetime volume) of the computation. At the algorithmic level, we can use more query-efficient algorithms that require a larger number of qubits. In addition, it may be possible to parallelize a calculation over multiple FTQCs---for example, sweeping over different parameters for the same calculation in parallel. At the logical level we can use additional logical qubits or increased non-locality of logical gates to realize subroutines with reduced circuit depth~\cite{Yuan2025ConnectivityConstraint,choi2011AdderConnectivity}. Moreover, we can parallelize sequential logical operations using gate teleportation~\cite{gottesman1999viabilityUniversalQC}. At the FTQC level, we can use additional qubits to increase the rate of magic state preparation for non-Clifford gates, or use QEC codes in 3D that can realize logical operations with reduced time overhead. At the physical level, there is often a time-accuracy tradeoff in physical gates and measurements, which could be converted into a spacetime tradeoff through quantum error correction. It is important to quantify the impact of these spacetime tradeoffs. In Sec.~\ref{Sec:FH_results} we present resource estimates that illustrate a selection of these tradeoffs by comparing different compilation schemes for a standard 2D planar architecture with lattice surgery, for the problem of simulating dynamics of the Fermi-Hubbard model.

The focus of this perspective is on strategies for optimizing logical clock speeds in the FTQC era. However, in the short term, we anticipate less focus on logical clock speeds, and so the strategies advocated for in this perspective will be less applicable. In the early QEC era, it will likely be acceptable to have slow logical clock speeds. Anticipated applications of early error-corrected devices center on scientific discovery~\cite{preskill2025megaquop}, targeting problems with limited classical competition. In these demonstrations of feasibility, scientists are likely to use every trick available to maximize logical computational volume, such as quantum error mitigation techniques~\cite{suzuki2022QEM,aharonov2025QEM} or more dense encodings~\cite{gidney2025yoked,kobori2025LoadStore}---even if these negatively impact logical clock speed. Nevertheless, these approaches will eventually give way to full fault tolerance, and we must look ahead to ensure that we do not engineer ourselves into a corner that limits us to only niche calculations that can tolerate slow clock speeds.

\section{Paradigms of fault-tolerant quantum computation}\label{Sec:FTQC}
Qubit modalities can be compared using many different metrics, including scalability, cost, control schemes, physical error rates (of gates, measurement, and idling), connectivity, and speed of physical operations. A well-designed FTQC architecture will exploit the capabilities of a candidate physical qubit to offset the qubit's shortcomings. Optimized FTQC architectures can then be compared by performing quantitative end-to-end resource estimates for specific applications.

We will focus on gate-based FTQC using surface and color codes in 2D and 3D. These codes have been extensively studied as realistic architectures, underpin many industrial roadmaps, and are currently the subject of experimental testing. As such, they provide a pragmatic benchmark against which more exotic proposals can be compared (for example, quantum low-density parity-check (LDPC) codes with high rate and code distance~\cite{Breuckmann2021}, which we briefly discuss in Sec.~\ref{Subsec:LDPC}). We first discuss the most well-studied paradigm, which we refer to as `standard 2D'. This consists of planar surface or color codes together with lattice surgery. This paradigm can be compared to proposals utilizing $2$D or $3$D codes, with transversal gates (which, by definition, do not couple physical qubits within the same code block). These proposals use additional hardware capabilities (e.g., non-local gates or qubit rearrangement) to realize more flexible FTQC schemes.

\subsection{Enabling universal fault-tolerant quantum computation}\label{Subsec:FTQC_requirements}

\begin{table*}[]
    \centering
    \begin{tabular}{c|c}
        \textbf{Timescale} & \textbf{Details} \\ \hline \hline
        Syndrome extraction (SE) round & $\mathcal{O}(1)$ physical gates, physical measurement. \\ \hline
        Reaction time ($\tau_r$) & Physical measurement, data transfer, decoding. \\ \hline
        Logical Clifford gate time ($\tau_c$) & \makecell{Lattice surgery: $\mathcal{O}(d)$ SE rounds.\\ Transversal gates: one layer of physical gates plus $\mathcal{O}(1)$ SE rounds.} \\ \hline
        \makecell{Logical non-Clifford gate time ($\tau_{nc}$)} & \makecell{Magic state teleportation: Logical Clifford time, plus reaction time, \\ plus conditional Clifford correction.} \\ \hline
        Single factory magic state preparation ($\tau_f$) & \makecell{Typically many SE rounds. \\ E.g., $18$--$470$ SE rounds~\cite{litinski2019magicstate} or $54$--$500$ SE rounds~\cite{gidney2024magiccultivation}. } \\ \hline
    \end{tabular}
    \caption{A summary of the timescales considered in our FTQC paradigms. Quantitative literature estimates for superconducting physical qubits and 2D surface code logical operations are shown in Fig.~\ref{Fig:SuperconductingTimescales}.}
    \label{tab:Timescales}
\end{table*}

Universal quantum computation can be realized with
the following ingredients: state preparation in $\ket{0}$, a universal gate set, and the ability to measure qubits in the $\{\ket{0}, \ket{1}\}$ basis. Implementations on bare physical qubits would not be reliable. Instead, FTQC based on QEC codes provides a route to reliable implementation of universal quantum computation. A high-level design principle for fault-tolerance based on QEC codes is to replace each logical qubit with a block of physical qubits encoded into a QEC code, and replace each logical operation by a fault-tolerant gadget acting on these physical qubits. In the absence of noise, the gadget perfectly implements the desired logical operation. When physical errors occur, the fault-tolerant gadget minimizes the spread of errors between physical qubits in the same code block. An example fault-tolerant gadget is a transversal logical gate for a QEC code. The fault-tolerant gadgets are interspersed with rounds of syndrome extraction, where all stabilizer generators of the QEC code are measured. We require sufficient rounds of syndrome extraction in the circuit to diagnose errors and capture their possible spread through the circuit. The frequency of syndrome extraction can be optimized by balancing the noise introduced by syndrome extraction with the noise introduced by the fault-tolerant gadgets. For example, if idling errors are low compared to two-qubit gate errors, then syndrome extraction may be performed less frequently for an idling logical qubit. In the paradigms we consider, the fault-tolerant gadgets required are realized using the following methods.
\begin{itemize}[itemsep=-1pt, topsep=0pt]
    \item State preparation in $\ket{0}$: All of the paradigms we consider can realize logical state preparation by preparing all physical qubits in $\ket{0}$, and measuring the stabilizers for the code.
    
    \item Universal logical gate set: The most widely studied universal logical gate set is single- and two-qubit Clifford + $T/$Toffoli. We remark
    that no quantum error correcting code supports a universal set of transversal gates~\cite{eastin2009RestrictionsTransversal}.
    \begin{itemize}[itemsep=-1pt, topsep=0pt]
        \item Logical Clifford gates: The paradigms we consider implement Clifford gates either directly via transversal gates, indirectly via logical Pauli measurements with lattice surgery, or implicitly via classical tracking in software using Pauli-based computation.
        \item Logical non-Clifford gates: The paradigms we consider typically implement logical non-Clifford gates by using a Clifford circuit to consume a specially prepared logical magic state. Magic state preparation can be treated as a black box. This provides a framework which encompasses magic state distillation~\cite{bravyi2005UniversalQC,litinski2019magicstate} and magic state cultivation~\cite{gidney2024magiccultivation}, as well as transversal implementations in 3D topological codes~\cite{beverland2021costUniversality}. It is typical to dedicate a fraction of the total physical qubits to magic state production---referred to as magic state factories (MSFs).
    \end{itemize}
    
    \item Logical Pauli measurements (with classical feed-forward): Reliably inferring logical measurement outcomes from a history of noisy syndrome measurements requires classical decoding algorithms and their hardware implementations. In many cases (for example, $T$-state teleportation) a logical measurement outcome determines the next logical quantum operation. The classical decoder must have sufficiently high throughput to keep pace with the rate of syndrome accumulation in the circuit (avoiding the backlog problem~\cite{terhal2015QECforQuantumMemories}), and should have low overall latency to ensure a high logical clock speed~\cite{Battistel2023RealTime}. This is formalized through the notion of `reaction time'.
\end{itemize}

We summarize the timescales of these operations in Table~\ref{tab:Timescales}.

\subsection{On the importance of the reaction time}\label{Subsubsec:ReactionTime}

\begin{figure}
    \centering
    \includegraphics[width=0.85\linewidth]{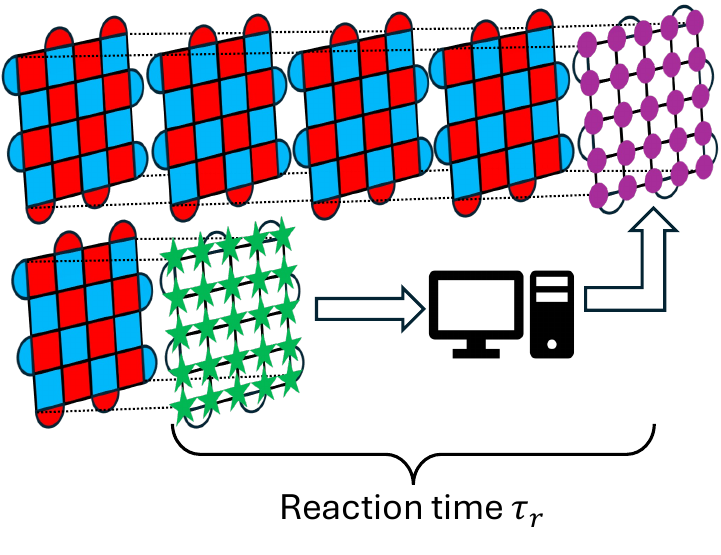}
    \caption{An illustration of the reaction time delay for a surface code computation. Time flows from left to right. Blue/red squares denote syndrome extraction. The physical qubits encoding the lower logical qubit are measured (green stars). The physical measurement outcomes are communicated to a classical processor for decoding, and decoded (using the syndrome history) to infer the logical measurement outcome. The logical measurement outcome determines the operation to be applied to the upper logical qubit. The instructions are then communicated back to the quantum processor, and implemented (purple circles). While the upper logical qubit waits for its next logical operation to be determined, syndrome extraction is carried out as usual.}
    \label{fig:Reaction_time}
\end{figure}

The reaction time of a given logical operation, $\tau_r$, is the time taken to collect classical information from a quantum system through physical qubit measurements, classically process the information, and initiate a quantum operation conditioned on the classical outcome. This is illustrated in Fig.~\ref{fig:Reaction_time} for a pair of logical qubits encoded in the surface code. The reaction time is system dependent, as it depends on the speed of physical qubit operations, classical processing, and latency of information transfer.

Although there exist measurement-free approaches to universal FTQC~\cite{webster2022UniversalFTQC,butt2024MeasurementFree,veroni2024MeasurementFree}, measurements are ubiquitous in the dominant FTQC paradigms. In particular, measurements are used for:
\begin{itemize}[itemsep=-1pt, topsep=0pt]
    \item Inferring logical errors by measuring error syndromes and decoding.
    \item Implementing logical operations, such as those implemented via lattice surgery.
    \item Teleporting quantum gates, such as in $T$-state teleportation.
    \item Code switching and gauge fixing~\cite{bombin2015gaugeColorCodes,Paetznick2013,Anderson2014}, as discussed in Sec.~\ref{sec:3D}.
    \item Optimizing quantum circuits with  mid-circuit logical measurements, such as in measurement-based uncomputation~\cite{gidney2018_halving_addition} or in the  semiclassical quantum Fourier transform \cite{griffiths1996semiclassicalQFT}.
\end{itemize}
In all of these cases, we must classically process the measurement outcomes to determine whether to apply a conditional operation. The reaction time induces a delay if, within the time taken to learn the measurement outcome, the circuit must \textit{physically change} in response to the measurement outcome. For example, the conditional $S$ gate arising from $T$-state teleportation cannot be neatly commuted through a subsequent Hadamard gate, and so there is a time overhead of $\tau_r$ to infer if the $S$ gate is required. In contrast, there is no delay from the reaction time if the effect of the conditional operation can be handled \textit{purely in software}. For example, a conditional logical Pauli operation can be commuted through a subsequent logical Clifford operation without changing the circuit---the conditional Pauli can be tracked through the Clifford in software. A consequence of this latter example is that deterministic Clifford circuits can be perfectly parallelized using gate teleportation, as any corrective Pauli operations can be commuted in software to the end of the circuit.

The `reaction depth' or `measurement depth' of a logical circuit is the minimum number of reaction time $\tau_r$ delays required to run the circuit. The reaction depth places a limit on our ability to parallelize quantum computation with gate teleportation, as the reactive operation must be separated in time by $\tau_r$ from its change to the circuit. As discussed above, teleporting in $T$ gates will introduce Clifford corrections (with probability one half). In reaction-limited quantum computation, we effectively teleport all gates. When teleporting in non-Clifford gates (e.g., $T$) we also \textit{selectively teleport} the associated corrective Clifford gate (e.g., $S$)~\cite{gidney2019flexible,litinski2019gameofsurfacecodes}. By selectively teleport, we mean that the application of the gate (up to Pauli corrections) can be decided at a later point in the circuit by the choice of measurement basis of the Clifford correction resource state. Pauli corrections from teleportations can be commuted through the circuit in software. Working chronologically through the circuit, one finds that the execution bottleneck becomes inferring the measurement basis of the Clifford correction resource states. Each Clifford correction depends on the measurement outcome of the previous Clifford correction qubit, and so sequential corrections are separated by time $\tau_r$. We provide a more detailed discussion in App.~\ref{App:time_optimal_computation}.

Because this approach saturates the reaction depth, it is known as time-optimal (or reaction-limited) quantum computing~\cite{fowler2012time}, where the claim of `time-optimal' is valid when $\tau_r$ is of a similar timescale to other physical operations.  
The run time of the circuit is given by the reaction depth multiplied by $\tau_r$. Notably, the run time of a circuit is determined only by its reaction depth, physical measurement speed (in the $Z$ and $X$ bases), and the speed of classical processing and communication---and is independent of other timescales, such as the two-qubit gate time. Reaction-limited computation contributes a space overhead proportional to $\tau_c/\tau_r$ in order to store the teleportation modules required.

\subsection{Standard 2D}\label{Subsec:Standard2D}

\begin{figure*}
    \centering
    \includegraphics[width=0.95\linewidth]{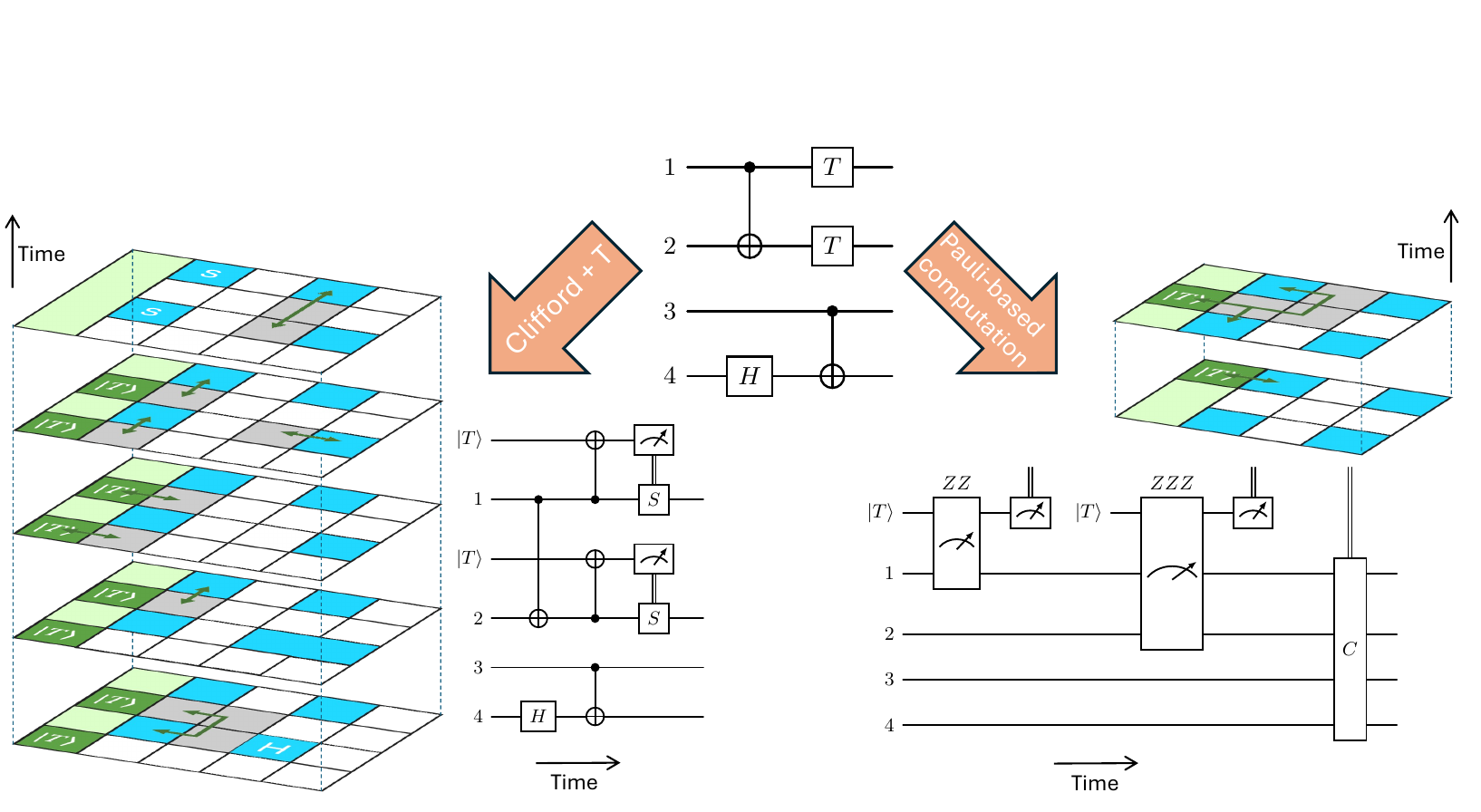}
    \caption{Circuit diagrams and logical timesteps for a circuit compiled to Clifford + $T$ (left), and to Pauli-based computation (PBC, right). In Clifford + $T$ compilation, logical Clifford gates are explicitly implemented via lattice surgery or transversal gates, and logical operations may be implemented in parallel if sufficient magic states and routing space are available. In PBC, Clifford gates are commuted through the circuit, leaving a sequence of non-Clifford Pauli-product-rotations, which can be implemented by performing (via lattice surgery) a logical Pauli measurement of the corresponding qubits and a magic state. The Clifford frame $C$ is tracked, and can be commuted through subsequent gates. This example highlights some of the benefits and drawbacks of PBC; Clifford gates do not have to be explicitly implemented, lowering the gate count of the circuit. However, the $T$ gates that were parallel in the original circuit are implemented sequentially after compiling to PBC. In the diagrams showing logical timesteps, time flows upwards. Logical data qubits are depicted by blue squares, magic state factories are light green regions (production of high fidelity magic states may require additional unshown patches), magic states $\ket{T} = T\ket{+}$ are dark green squares, idle routing space is denoted by white squares, and routing space that is used in each time step by lattice surgery is depicted as gray squares. 
    Green arrows indicate qubits that are involved in a logical Pauli measurement performed via lattice surgery. `H' and `S' are used to denote latice surgery implementations of the Hadamard and $S$ gates, respectively. 
    }
    \label{fig:CliffordT_PBC}
\end{figure*}

The first paradigm we consider consists of planar QEC codes, such as the (rotated) surface code, or 2D color code. Logical gates are implemented using lattice surgery, which involves changing the stabilizers measured during syndrome extraction~\cite{horsman2012latticeSurgery}. In addition to performing logical operations between adjacent logical qubits, lattice surgery can be used to implement long-range operations, provided that there is an uninterrupted path of routing space (physical qubits currently not being used in the computation) between the target boundaries of the logical qubits. Below, we will compare the two main compilation schemes for the standard 2D paradigm, illustrated in Fig.~\ref{fig:CliffordT_PBC}. This paradigm has been extensively studied and has a number of strengths: it is compatible with an architecture with nearest-neighbor qubit connectivity in 2D, has high threshold error rates, has a number of logical gate schemes, and can be realized in modular architectures with minimal reduction in performance~\cite{Ramette2024}. A drawback of this paradigm is that 2D topological codes have poor scaling of code parameters compared to optimal constructions~\cite{Bravyi2009,Bravyi2010}. In addition, implementing gates via lattice surgery incurs a time overhead of $\mathcal{O}(d)$ syndrome extraction rounds, where $d$ is the code distance, in order to suppress logical failures arising from physical measurement errors. For reference, existing surface code resource estimates for end-to-end applications typically require $d = 25$ or larger (assuming physical error rates of $p=10^{-3}$~\cite{fowler2012SurfaceCodes}). We note that by using the aforementioned methods of time-optimal/reaction-limited quantum computing~\cite{fowler2012time} (discussed in more detail in App.~\ref{App:time_optimal_computation}), it is possible to realize consecutive operations with an effective time cost of $\tau_r$ despite using lattice surgery.

Magic state preparation has been optimized within the standard 2D paradigm. We will restrict our discussion to preparation of the $T$-states here. $T$-states with infidelity comparable to the physical noise rate can be injected into the surface code using physical circuits with 2D local connectivity~\cite{li2015magic,lao2022MagicState}. Multiple low-fidelity $T$-states can be distilled into a $T$-state of higher fidelity using logical Clifford circuits, implemented using lattice surgery. Using Pauli-based computation (discussed below), these circuits can be recompiled to provide spacetime tradeoffs~\cite{litinski2019gameofsurfacecodes}. For example, the circuit for 15-to-1 distillation can be re-expressed to use 5 logical qubits and a depth of 11 lattice surgery operations~\cite{litinski2019gameofsurfacecodes} (11 logical qubits including routing space). 2D color codes have also been investigated for magic state preparation~\cite{lee2025ColorCodeDistillation}. In particular, the ability to transversally perform controlled measurements of single-qubit logical Clifford gates in 2D color codes has been used to optimize the cost of preparing encoded magic states with high fidelity via magic state cultivation~\cite{chamberland2019faultTolerantMagicStateFlagQubits,itogawa2025ZeroLevelDistillation,gidney2024magiccultivation}.

\subsubsection{Clifford + $T$/Toffoli}
The first compilation scheme is to decompose a quantum circuit in terms of logical Clifford + $T$/Toffoli gates, and then to implement all logical gates explicitly using physical operations that are geometrically local in 2D. For the surface code, single-qubit logical Clifford gates such as Hadamard and $S$, and two-qubit logical Clifford gates such as CNOT can be realized with a planar layout via lattice surgery~\cite{horsman2012latticeSurgery,fowler2018}. The 2D color code has the added benefit of transversal single-qubit logical Cliffords, such as Hadamard and $S$. Lattice surgery can implement a non-local logical CNOT, or even a multitarget CNOT, in $\mathcal{O}(d)$ rounds, provided there is routing space between appropriate boundaries of the control and target qubit patches~\cite{fowler2018}. This multitarget/FANOUT CNOT gate can be particularly useful for reducing the depth of common quantum circuits, including quantum lookup-tables (Sec.~\ref{Subsec:LookupTablesQROM} and~\cite{babbush2018EncodingElectronicSpectraLinearT,gidney2019windowedQuantumArithmetic,lee2021EvenMoreEfficientChemistryTensorHyp}), and the SWAPUP$^*$ operation (App.~\ref{Subsec:FH_SELECT} and~\cite{low2018tradingTgatesforDirtyQubits,Wan2021exponentiallyfaster}). In general, it may be possible to reduce compilation inefficiencies and save resources by bypassing the compilation of algorithmic subroutines to unitary gates, and compiling directly to native lattice surgery operations~\cite{deBeaudrap2020ZXcalculusLatticeSurgery}.

The main benefit of the Clifford + $T$/Toffoli scheme is that logical gates can be realized in parallel, provided that their constituent lattice surgery operations act on disjoint logical qubits and routing space. Constraining the logical computation to 2D leads to a routing problem that is NP-hard to solve exactly~\cite{herr2017NPhard}. Nevertheless, a number of works have investigated using heuristic algorithms to solve the routing problem, in some cases also optimizing logical qubit layouts and the number of MSFs~\cite{beverland2022SurfaceCodeCompilation,tan2024satCompiler,molavi2023compilation,watkins2024Compiler,leblond2024realisticCompiler,hamada2024EfficientLatticeSurgeryRouting,herzog2025latticesurgerycompilationsurface}. A notable observation is that placing MSFs at the boundary of the computational region leads to a maximum magic state consumption rate that is proportional to the perimeter of the computational region, limiting the number of $T$ gates that can be applied in parallel~\cite{beverland2022SurfaceCodeCompilation}. A possible solution to this issue, which we will assess in our resource estimates, is to locate the MSFs within the computational area, at a cost of increasing the routing space (see also Ref.~\cite{hirano2025localityPBC}). There have also been hardware proposals to mitigate the routing problem by using a bi-layer architecture, at a cost of increasing the complexity of engineering~\cite{ueno2024bilayerCompiler,ruiz2025ldpcCAT}.

\subsubsection{Pauli-based computation}
Pauli-based computation (PBC) is a common compilation strategy for the standard 2D architecture, popularized by Ref.~\cite{litinski2019gameofsurfacecodes}. PBC works by re-writing the Clifford + $T$ gate set as Pauli-product-rotations (PPRs) of the form $\exp(i\theta P)$ for $P$ Pauli and $\theta \in \{\pi/2, \pi/4, \pi/8\}$, and then commuting Pauli ($\theta = \pi/2$) and Clifford ($\theta = \pi/4$) PPRs through non-Clifford ($\theta = \pi/8$) PPRs, to the end of the circuit. This causes the non-Clifford PPR to `spread', gaining support on multiple logical qubits. The circuit can then be expressed as a sequence of multiqubit non-Clifford PPRs. Each logical multiqubit non-Clifford PPR can be implemented by performing (via lattice surgery) a logical Pauli product measurement (PPM) of the corresponding qubits and a $T$-magic state, distilled in an MSF. The benefit of PBC is that it eliminates Clifford gates from the circuit. This reduces the total gate count, and simplifies resource estimates to counting the total number of $T$ gates in the circuit. This makes it easy to do resource estimates with general-purpose architectures, such as those introduced in Ref.~\cite{litinski2019gameofsurfacecodes}. Each PPR is completed in $\mathcal{O}(d)$ rounds of syndrome extraction, with schemes to reduce the constant prefactor~\cite{chamberland2022TwistFree, Prabhu2022Tels}. The scheme is also compatible with reaction-limited computation, at a cost of high space overhead~\cite{litinski2019gameofsurfacecodes}.

A number of works have investigated compiling algorithms via PBC~\cite{litinski2019gameofsurfacecodes,Beverland2022Requirements,silva2024multi}. The main drawback of PBC is that the support of $T$ gates will quickly start to overlap. This reduces parallelism in the circuit, potentially increasing circuit depth. A secondary issue of the growth in support is the need to decode multiqubit lattice surgery operations that potentially have support on all logical qubits in the algorithm.

\subsubsection{Natural hardware: superconducting qubits}
The standard 2D paradigm is well suited for superconducting qubits~\cite{Blais2004,devoret2013superconductingCircuits,blais2021circuitQuantumElectrodynamics}. Most prominently, 2D planar codes do not require long-range connectivity, which is more challenging to engineer in superconducting qubit systems. In addition, the surface code has a high threshold error rate that is compatible with the error rates of state-of-the-art superconducting qubits~\cite{google2025Below}, and it can tolerate higher noise rates across interconnects in modular architectures~\cite{Ramette2024}. To their credit, superconducting qubits have relatively fast physical clock speeds (approximately \SI{1}{\micro \second} per syndrome extraction round) which offsets the $\mathcal{O}(d)$ time overhead of lattice surgery operations. The superconducting qubit standard 2D architecture has been extensively studied, resulting in a number of physical resource estimates for various applications~\cite{babbush2018EncodingElectronicSpectraLinearT,lee2021EvenMoreEfficientChemistryTensorHyp,gidney2021HowToFactor,Beverland2022Requirements}. A promising emerging direction is to engineer noise bias in superconducting qubits, which can reduce the overheads of error correction~\cite{kubica2022erasure,Teoh2023,levine2024DualRail,Gu2023fault-tolerant,Guillaud2019,Puri2020,Putterman2025Cat} 
and magic state preparation~\cite{chamberland2022buildingFTQC,singh2022BiasedMSD,gouzien2023catCodeEllipticCurve,vaknin2025magic,jacoby2025magic}, leading to improved resource estimates for applications~\cite{gouzien2023catCodeEllipticCurve,chamberland2022buildingFTQC}. In Fig.~\ref{Fig:SuperconductingTimescales}, we show a representative set of currently achievable timescales for superconducting qubits, along with the implied logical gate times. It is important to note that, in general, these times have not been optimized for speed, but rather for error rates and overall system engineering feasibility in the context of basic QEC experiments \cite{bengtsson2024modelsOfReadout}.

\begin{figure*}
\centering
\begin{tikzpicture}

\definecolor{deepurple}{RGB}{88, 24, 69}
\definecolor{goldyellow}{RGB}{255, 190, 11}
\definecolor{mutedpurple}{RGB}{146, 111, 159}
\definecolor{warmyellow}{RGB}{255, 200, 87}
\definecolor{brightpurple}{RGB}{128, 0, 128}  
\definecolor{brightyellow}{RGB}{255, 223, 0}  

\def\textSize{\normalsize}
\def\labelSize{\footnotesize}

\def\tikzVspace{0.56cm}        
\def\tikzVlinespace{2.3cm}    
\def\tikzTextSep{0.1cm}     
\def\tikzTextcolwidth{7.8cm}  
\def\tikzNumvlines{4}       
\def\tikzTopspace{0.2cm}    
\def\tikzHoffset{1.3cm}     
\def\tikzHoffsetRight{1.4cm}     
\def\tikzCircleLabelOffset{0.1cm}     
\def\tikzCircleRadius{0.15cm}   
\def\numRows{6}   

\def\tikzHlinewidth{\tikzHoffset+\tikzHoffsetRight+\tikzNumvlines*\tikzVlinespace-\tikzVlinespace}    

\def\xShift{0 cm}

\def\vlabels{%
    {\SI{100}{\nano \second}},%
    {\SI{1}{\micro \second}},%
    {\SI{10}{\micro \second}},%
    {\SI{100}{\micro \second}}%
}

\foreach [count=\i starting from 1] \label in \vlabels {
    \pgfmathsetmacro{\x}{(\i-1)}
    \draw[dotted] (\xShift+\tikzHoffset+\x*\tikzVlinespace,\tikzTopspace) -- 
        (\xShift+\tikzHoffset+\x*\tikzVlinespace,{-(\numRows-1)*\tikzVspace-\tikzTopspace});
    \node[anchor=south,rotate=0] at (\xShift+\tikzHoffset+\x*\tikzVlinespace,\tikzTopspace+0.1cm) {\textSize \label};
}

\foreach [count=\y starting from 0] \colorVal/\x/\len/\pow/\circlelabel in {%
    mutedpurple/{Time for two-qubit physical quantum gate}/4.2/1/{\SI{42}{\nano \second} \cite{google2025Below}},%
    mutedpurple/{Physical measurement time}/5/2/{\SI{500}{\nano \second} \cite{google2025Below,bengtsson2024modelsOfReadout}},%
    mutedpurple/{Time for SE round}/1.1/3/{\SI{1.1}{\micro \second} \cite{google2025Below}},%
    mutedpurple/{Estimated time $\tau_c$ for logical Clifford gate}/2.8/4/{\SI{28}{\micro \second}},%
    mutedpurple/{Estimated time $\tau_{nc}$ for logical non-Clifford gate}/5.5/4/{\SI{55}{\micro \second}},%
    mutedpurple/{Estimated time $\tau_f$ to produce one magic state}/9.1/4/{\SI{91}{\micro \second} \cite{litinski2019magicstate}}%
    } {
    \node[anchor=east, text width=\tikzTextcolwidth, text ragged left] 
        at (\xShift-\tikzTextSep,-\y*\tikzVspace) {\textSize \x};
    
    \draw[black] (\xShift-\tikzTextcolwidth,-\y*\tikzVspace + \tikzVspace/2) -- (\xShift+\tikzHlinewidth,-\y * \tikzVspace + \tikzVspace/2);
    
    \pgfmathsetmacro{\logvalueAdjusted}{(-2+\pow+log10(\len))*\tikzVlinespace}
    \draw[fill=\colorVal] (\xShift+\tikzHoffset+\logvalueAdjusted,-\y*\tikzVspace) circle (\tikzCircleRadius);
    
    \node[anchor=west, font=\scriptsize] at (\xShift+\tikzHoffset+\logvalueAdjusted+\tikzCircleLabelOffset,-\y*\tikzVspace) {\labelSize \circlelabel};

}
\draw[black] (\xShift-\tikzTextcolwidth,-\numRows*\tikzVspace + \tikzVspace/2) -- (\xShift+\tikzHlinewidth,-\numRows * \tikzVspace + \tikzVspace/2);

\end{tikzpicture}


\caption{\label{Fig:SuperconductingTimescales} Selection of currently achievable time scales relevant for superconducting quantum processors in the `standard 2D' FTQC paradigm. Physical quantum gate and measurement times are taken from the 105-qubit processor reported in \cite{google2025Below}, which was used to demonstrate elements of QEC in the standard 2D paradigm for surface codes up to $d=7$.  Logical gate times assume a lattice surgery approach requiring $d$ SE rounds (\SI{1.1}{\micro \second} per SE round), with an extra factor of 2 for non-Clifford gates to perform the conditional correction. Magic state production time varies widely with the chosen MSF strategy, the target output error rate, and the physical error rate. At $p=0.1\%$ physical error rate, one MSF can produce one $\ket{T}$ magic state with output error probability $2.7 \times 10^{-12}$ in 83 SE rounds \cite[Table 1]{litinski2019magicstate}, suitable for some applications listed in \cref{tab:ResourceEstimates}. 
}

\end{figure*}

\subsection{Alternatives to standard 2D}

\subsubsection{2D with transversal gates}
\label{sec:2Dtrans}
In this paradigm, we consider a platform realizing 2D planar codes augmented with non-local transversal logical gates. For example, the surface code admits transversal CNOT, $H$ (up to a patch rotation), and $S$ (with respect to a particular qubit partition~\cite{kubica2015unfoldingColorCode,moussa2016transversalCliffordGates}). This paradigm encompasses recent proposals for algorithmic fault-tolerance~\cite{Zhou2024AlgorithmicFT}\footnote{For a discussion of the prospects of combining PBC with algorithmic fault tolerance, with the aim of reducing the time per PPR below $\mathcal{O}(d)$ rounds, we refer the reader to \cite[Appendix VI]{Zhou2024AlgorithmicFT}.} and active volume compilation~\cite{litinski2022activeVolume}. \\

Transversal logical entangling gates can be used to reduce the number of syndrome extraction rounds after each logical gate from $\mathcal{O}(d)$ to $\mathcal{O}(1)$, where $d$ is the code distance. For example, one can make fault-tolerant syndrome measurements by preparing special ancilla qubit states, incurring large qubit overhead~\cite{steane1997ActiveStabilization,knill2005QCwithNoisyDevices}. Algorithmic fault-tolerance (AFT) achieves a similar run time improvement, without the large qubit overhead. This is achieved by tracking the deterministic propagation of Pauli errors through transversal Clifford gates, which limit the spread of errors. Logical non-Clifford gates, such as $T$, are implemented by teleporting in distilled magic states using transversal logical Clifford operations.
The crucial component of AFT is correlated decoding~\cite{Cain2024}, which uses the syndrome information across the entire spacetime volume of the logical circuit.
Initial attempts to use efficient decoders work only in limited cases~\cite{beverland2021costUniversality,sahay2025TransversalCNOT,wan2025iterativetransversalcnotdecoder}. However, recent work has shown how to decode AFT using computationally efficient decoders based on matching~\cite{cain2025FastCorrelatedDecoding,serra2025DecodingTransversal}. These approaches rely on the observation that one can individually decode each logical observable by separately backpropagating them through the Clifford circuit and processing the syndrome information only from the relevant past spacetime volume. Although it is formally efficient to decode the resulting syndrome history using matching, it may be too slow to enable decoding with sufficiently low latency, which in turn motivates recent work to modularize the decoding of AFT~\cite{serra2025DecodingTransversal, turner2025ScalableTransversalDecoding}. The main benefit of the AFT scheme is its ability to reduce the spacetime volume and wall time of an algorithm by a factor of $\mathcal{O}(d)$, in contrast to alternative single-shot approaches, which typically trade reduced time for extra space. The main drawback is the increased decoding difficulty, compared to that of decoding surface code memory~\cite{Skoric2023parallel}, or computation via lattice surgery~\cite{Bombin2023modular}. Improving the practicality of decoding AFT and understanding observed pathological behavior of decoders in edge cases are at the frontier of practical QEC research~\cite{cain2025FastCorrelatedDecoding,serra2025DecodingTransversal,turner2025ScalableTransversalDecoding}.\\

The active volume (AV) scheme was proposed in Ref.~\cite{litinski2022activeVolume} as a method of reducing the spacetime volumes of computations. The scheme uses a specified connectivity; transversal logical gates between qubits $i$ and $j$ are implementable if $|i-j|\leq \mathcal{O}(1)$, and transversal logical SWAP operations between $i$ and $j$ are available if $|i-j| = 2^k$ for $k\in [0,\lfloor\log(N)\rfloor]$, where $N$ is the number of logical qubits. This connectivity, resembling that of a hypercubic lattice, enables rapid reshuffling of the logical qubit positions, and fast initialization of Bell pairs and Bell basis measurements. These ingredients facilitate parallelization of sequential logical operations via gate teleportation. The aim of the scheme is to minimize the circuit depth, given a restricted logical footprint. The AV approach uses Clifford + $T$/Toffoli compilation, rather than PBC in order to maintain and exploit parallelism in the circuit. The AV scheme has been shown to provide run time reductions of 10--100$\times$ for certain problems in cryptanalysis~\cite{litinski2022activeVolume,litinski2023EllipticCurvesBaseline,garn2025ActiveVolumeECC} and quantum chemistry~\cite{caesura2025ActiveVolumeChemistry}. The main limitation of the method is its connectivity requirements, which may be infeasible for many hardware platforms, and increase the engineering difficulty in others.

Transversal logical entangling gates and logical reconfigurability of qubits encoded in 2D topological codes also offer benefits for the preparation of magic states. Transversal logical entangling operations have recently enabled experimental demonstration of 5-to-1 distillation of magic states encoded in 2D color codes, in a neutral atom processor~\cite{sales2025experimentalMSD}. On the theoretical front, AFT has been observed as being well suited for magic state distillation circuits, due to their limited circuit depth~\cite{Zhou2024AlgorithmicFT,cain2025FastCorrelatedDecoding}, which can be further reduced using transversal logical operations and logical reconfigurability~\cite{fazio2025LowTransversalMagic}. Alternatively, non-local transversal gates can be used to reduce the costs of magic state cultivation of $T$-states in the color code, or to broaden the reach of the technique to $CCZ$-states in the surface code~\cite{vaknin2025magic,chen2025EfficientCultivation,sahay2025foldCultivation,claes2025cultivating}.

\subsubsection{3D}\label{sec:3D}

In this architecture, we assume that some part or the entire quantum processor consists of a 3D arrangement of qubits that supports geometrically local quantum operations.
This spatial structure enables native implementation of quantum codes such as the 3D gauge color code~\cite{bombin2015gaugeColorCodes} and the 3D subsystem toric code~\cite{kubica2022singleShotQECtoric,bridgeman2023liftingTopologicalCodes}.
The primary benefits of 3D architectures include the ability to implement transversal non-Clifford gates and to perform single-shot QEC.
These two features impact the time overhead of fault-tolerant quantum computation.

Transversal gates, or more generally locality-preserving quantum operations, are especially desirable because they are simple to implement and naturally fault-tolerant.
Unlike 2D quantum codes~\cite{bravyi2013,Beverland2016}, 3D quantum codes can admit transversal implementation of non-Clifford gates~\cite{Bombin2013,kubica2015unfoldingColorCode}, which in turn are essential for universal quantum computation.
Consequently, one can achieve universality by code switching~\cite{Paetznick2013,Anderson2014,bombin2015gaugeColorCodes,kubica2015universalTransversalGates} between two codes: a code with a transversal non-Clifford gate, such as the 3D stabilizer color code, and a code with transversal Clifford gates, such as the 3D gauge color code or even the 2D color code. We remark that logical entangling gates may be implemented as a product of physical entangling gates between corresponding qubits on the 2D boundaries of two 3D code blocks, as exemplified by a logical $CS$ gate in the 3D color code~\cite{bombin2018transversalgateserrorpropagation}. Also note that a similar discussion is also applicable to the family of toric codes.

As discussed in previous sections, QEC must be made robust to measurement errors during syndrome extraction, typically by repeating measurements and using an extended syndrome history, which leads to a significant time overhead~\cite{dennis2002TopologicalQuantumMemory,shor1996FTQC}. To mitigate this downside, one may seek quantum codes that facilitate single-shot QEC~\cite{bombin2015SingleShotFTQEC,campbell2019theorySingleShot}, that is, reliable QEC that is possible even in the presence of measurement errors, without the need for repeated measurements or an extended syndrome history. Notably, 3D architectures are sufficient to realize such quantum codes, including the 3D gauge color code and the 3D subsystem toric code.
While certain high-rate quantum LDPC codes may also exhibit this property~\cite{fawzi2018ConstantOverheadExpanderCode,breuckmann2022,gu2024}, they generally cannot be realized with architectures in three or fewer spatial dimensions with geometrically local gates.

In order to realize fault-tolerant quantum computation with 3D architectures, one can rely on 3D quantum codes described above that simultaneously have transversal gates and facilitate single-shot QEC.
The key advantage of this approach is that each basic operation at the logical level, including state preparation, QEC, code switching and both Clifford and non-Clifford gates, can be implemented using constant-depth quantum circuits composed of geometrically-local physical gates.
Consequently, these basic operations are executed in constant time (proportional to the depth of the corresponding quantum circuits), provided sufficiently fast processing of global classical information.
More precisely, both code switching and single-shot QEC are limited by the reaction time required by classical computation to find, respectively, a gauge fixing or an error recovery.
Thus, for the approach based on 3D quantum codes to be practical, the reaction time should not significantly exceed the implementation time of physical quantum operations.
Additionally, one must consider the increased qubit overhead typically associated with 3D quantum codes compared to their 2D counterparts.

An initial study~\cite{beverland2021costUniversality} suggested that code switching may not offer substantial overhead reductions and could even perform worse than magic state distillation, while still leaving room for future improvements in code switching protocols.
Importantly, the analysis highlighted the influence of various simplifying assumptions and approximations on the overhead estimates, as well as the critical role of color code decoding and its performance.
With single-shot state preparation achievable in 3D architectures, the resulting spacetime overhead of code switching can scale as $d^3$, where $d$ is the distance of the output code, either matching or even improving upon that of magic state distillation.
More recent work~\cite{Daguerre2025} has revisited this question in the context of small-scale implementations using distance-3 codes, with encouraging results.
These advances leverage techniques such as flag-based postselection and transversal gates, and have paved the way for a recent experimental demonstration of code switching between the 15-qubit and 7-qubit color codes~\cite{daguerre2025experimentaldemonstrationhighfidelitylogical}.
The recent progress indicates that despite the greater engineering challenges and higher qubit overhead, the potential for faster fault-tolerant quantum computation may well justify the associated costs inherent to 3D architectures.

\subsubsection{Natural hardware: neutral atoms}
Neutral atoms are a promising hardware platform, and have demonstrated logical reconfigurability that enables the implementation of non-local transversal gates~\cite{Bluvstein2023}. These capabilities could be used to realize the alternative paradigms discussed above---and in fact motivated the AFT scheme (as well as other specific proposals for quantum LDPC architectures~\cite{xu2024LDPC}). The AFT scheme provides a method to mitigate the comparatively slow physical clock speeds of neutral atoms. Underpinning a neutral atom realization of any of these alternative paradigms is the use of atom shuttling via optical tweezers to enable non-local gates. Following Ref.~\cite{zhou2025resourceAnalysisLowOverheadTransversal} and references therein, current shuttling times are determined by an effective acceleration of atoms of \SI[per-mode=symbol]{5500}{\metre\per\second\squared} and a typical separation between them of \SI{12}{\micro \metre}.
Assuming a $10 \times 10$ grid of $d=20$ surface code patches, this would yield an estimate \SIrange[range-phrase=--]{0.42}{1.57}{\milli\second} of time needed for a transversal CNOT gate (depending on the relative position of patches and assuming constant effective acceleration/deceleration during the first/second half of the atom trajectory).
Other typical parameters for neutral atom systems include
physical gate time (\SI{1}{\micro \second}) and measurement time (\SI{500}{\micro \second}).
These estimates are 10--100$\times$ higher than those for superconducting qubits with logical gates implemented via lattice surgery, highlighting the need for a more detailed, application-specific comparison.
There is a broad consensus that the performance of neutral atom systems can be still significantly improved.
Moreover, the impact of slower run times can be partially mitigated by AFT. For example, the neutral atom factoring resource estimate (5.6 days) in Ref.~\cite{zhou2025resourceAnalysisLowOverheadTransversal} is only 18$\times$ slower than the superconducting qubit estimate (7.3 hours) of Ref.~\cite{gidney2021HowToFactor} (both requiring 20 million physical qubits), despite the fact that the syndrome extraction time is estimated to be 900$\times$ longer. This relative $50\times$ saving is obtained by using AFT over lattice surgery, and by upgrading to lower-cost MSFs incorporating magic state cultivation~\cite{gidney2024magiccultivation}.

\subsection{Comments on progress on quantum LDPC codes}\label{Subsec:LDPC}
There has recently been rapid progress in the development of quantum LDPC codes~\cite{Breuckmann2021}, culminating in the construction of families of codes with asymptotically good parameters, that is, constant encoding rates and relative distances~\cite{breuckmann2021balancedProductQuantumCodes,panteleev2022GoodLDPC,leverrier2022QuantumTannerCodes,dinur2022goodquantumldpccodes}.
These codes necessarily require non-local interactions, and so it is reasonable to assume they will be more challenging to realize than low dimensional topological codes. While good quantum LDPC codes in theory excel as candidates for building a logical quantum memory, the practical performance of finite-size instances as an FTQC architecture is still under active investigation. There seems to be a tension between parallelizing logical operations and maximizing the logical encoding rate. Building on a generalization of lattice surgery techniques~\cite{Cohen2022low}, there have been recent proposals for heterogeneous FTQC architectures, comprising a quantum LDPC memory block, a surface code compute region, and an ancilla bus for logical information transfer~\cite{xu2024LDPC,stein2024architectures}. The clock speed is determined by the time to load logical qubits from the memory, and the size of the compute region. These approaches are being superseded by proposals which natively compute in multiple quantum LDPC blocks, with the aid of specialized ancilla patches (see Sec.~3.2 in Ref.~\cite{he2025extractorsqldpcarchitecturesefficient} for a comprehensive overview of recent progress and relevant references).

Quantum LDPC codes are a very active area of research, whose direction is hard to predict. Their immediate impact would appear to increase wall time of computations, while reducing qubit requirements. This is corroborated by recent resource estimates for a proposed quantum LDPC architecture~\cite{yoder2025tour}. Given the rapidly developing nature of this subfield and our focus on strategies for reducing wall time, we restrict our discussion to more well-established paradigms, which provide a benchmark against which future quantum LDPC-based proposals can be compared.

\section{Logical parallelism}
Quantum subroutines can have multiple equivalent circuit instantiations. Parallel implementations reduce gate depth (or depth of a particular type of gate, such as $T$/Toffoli), typically at a cost of more gates, increased range of entangling gates between logical qubits, and additional ancilla qubits. This can improve logical clock speeds, but the corresponding space overhead depends on the underlying FTQC approach considered. Of particular importance is the need to have a sufficient number of MSFs and routing space to saturate the rate of magic state consumption in the circuit. An additional challenge is managing resources when the rate of magic state consumption varies throughout the algorithm.

To date, the spacetime volume of MSFs and the desire to maintain modest physical qubit counts in resource estimates have led to a preference for compilations with minimal $T$/Toffoli-count, and sequential gate implementation.
The run time is then dominated by waiting for and consuming the magic states. Below we discuss explicit examples of circuits with both sequential and parallelized compilations, to illustrate the tradeoffs that must be considered when compiling quantum algorithms. Our chosen subroutines, quantum adders and quantum lookup tables, underpin compilations of many quantum algorithms, including those for cryptography~\cite{gidney2021HowToFactor,litinski2023EllipticCurvesBaseline}, chemistry~\cite{babbush2018EncodingElectronicSpectraLinearT,Berry2019QubitizationOfArbitraryBasisChemistry,lee2021EvenMoreEfficientChemistryTensorHyp,su2021FaultTolerantChemistryFirstQuantized}, optimization~\cite{sanders2020FTQCforCombOpt}, finance~\cite{chakrabarti2021threshold}, and dynamical quantum simulations of lattice gauge theories \cite{kan2021lattice, davoudi2023GeneralNonAbelian}.

\subsection{Quantum adders}

Quantum addition has a number of proposed implementations of varying quantum circuit depths. Generally, the goal is to coherently compute $\ket{x}\ket{y}\ket{0}\mapsto \ket{x}\ket{y}\ket{x+y}$ (out-of-place) or $\ket{x}\ket{y} \mapsto \ket{x}\ket{x+y \bmod 2^n}$ (in-place), where $x$ and $y$ are $n$-bit integers, possibly with the assistance of additional ancilla qubits.
The quantum adder with minimal Toffoli count is the ripple-carry quantum adder \cite{vedral1996rippleCarryAdder, cuccaro2004newquantumripplecarryaddition, gidney2018_halving_addition}, which computes the output bits sequentially, starting with the least significant bit and working toward the most significant, incurring depth linear in $n$. A number of methods exist for reducing the depth, including carry-lookahead \cite{thapliyal2020tcountqubitoptimizedquantum,wang2024optimaltoffolidepthquantumadder}, block-lookahead \cite{gidney2020QuantumBlockLookahead}, carry-save \cite{gossett1998quantumcarrysavearithmetic}, oblivious carry runways (approximate) \cite{gidney2019approximateEncodedPermutationsPiecewise}, quantum Fourier transform (approximate) \cite{draper2000additionQuantumComputer}, and direct re-compilation of the ripple-carry adder using conditionally clean ancillas \cite{remaud2025ancillafreequantumaddersublinear}. In fact, if errors of size $1/\mathrm{poly}(n)$ are tolerable, then $\mathcal{O}(\log\log(n))$  depth is asymptotically achievable \cite{gidney2019approximateEncodedPermutationsPiecewise}. Generally, reduced depth comes at the cost of larger Toffoli count, larger number of ancilla qubits, or both; some of the available choices for in-place addition are presented in Table~\ref{tab:quantum_addition}, and we refer the reader to Refs.~\cite{gidney2020QuantumBlockLookahead, orts2020reviewReversibleQuantumAdders,paler2022AnalysisArithmetic, wang2025comprehensiveStudyArithmetic} for more detailed comparisons. 

\begin{table}[ht!]
    \centering
    \begin{tabular}{c|c|c|c}
    \textbf{Method} & \textbf{Count} & \textbf{Depth} & \textbf{Ancillas} \\
    \hline
    \hline
    Ripple-carry \cite{cuccaro2004newquantumripplecarryaddition} & $2n$ & $2n$ & 1\\
    Ripple-carry \cite{takahashi2010quantumAdditionUnboundedFanout} & $2n$ & $2n$ & $0$ \\
    Ripple-carry \cite{gidney2018_halving_addition} & $n$ & $2n$ & $n$ \\
    \hline 
    Carry-lookahead \cite{thapliyal2020tcountqubitoptimizedquantum} & $7n$ & $ 4 \lg(n) $ & $2n$ \\
    Block-lookahead \cite{gidney2020QuantumBlockLookahead} & $5n-4b+8\frac{n}{b}$ & $ 6b + 4 \lg(\frac{n}{b}) $ & $2n+3\frac{n}{b}$ \\
    \hline
    \makecell{Conditionally clean \\ ancillas \cite{remaud2025ancillafreequantumaddersublinear}} & $\mathcal{O}(n \log(n))$ & $\mathcal{O}(\log^2(n))$ & $0$ \\
    \hline
    \makecell{Ripple-carry  + obliv.\\ carry runways \cite{gidney2019approximateEncodedPermutationsPiecewise}}  & ${\textstyle 2n + r\lg(\frac{r^2}{\epsilon^4})}$&  ${\scriptstyle \frac{2n}{(r+1)} + \lg(\frac{r^2}{\epsilon^4})}$ &  ${\textstyle r\lg(\frac{r}{\epsilon^2})}$  \\ \hline
\end{tabular}
\caption{The Toffoli count, reaction depth (where a parallel layer of Toffoli gates are assumed to be implemented via magic state teleportation with reaction depth-1),
and number of ancilla qubits required for a subset of quantum in-place addition strategies. Constant prefactors are reported when known, but only the asymptotically leading term in $n$ is included for compactness. Notation $\lg$ signifies $\log_2$. The parameter $b \in [1,n]$ is the blocksize for the block-lookahead adder, the parameter $r \in [1,\mathcal{O}(n/\lg(n))]$ is the number of oblivious carry runways, and the parameter $\epsilon$ is the error in the case of approximate addition (the runway length needed to achieve error $\epsilon$ is $\lg(r/\epsilon^2) + O(1)$ bits).
\label{tab:quantum_addition}}
\end{table}

The tradeoffs available lead to considerable nuance when compiling algorithms requiring quantum arithmetic. Quantum addition also features as a surprising ingredient in some important quantum primitives unrelated to arithmetic; for example, addition induces phase kickback $\ket{x} \mapsto e^{2 \pi i x/2^n}\ket{x}$ when performed into a Fourier ancilla state, efficiently passing information from the computational basis state into the phase of that state \cite{gidney2018_halving_addition, low2018tradingTgatesforDirtyQubits}.

By parallelizing the Toffoli gates, the shallower adders achieve lower  asymptotic spacetime volume in these applications, compared to the ripple-carry adder. However, they require a greater number of MSFs to match their larger consumption rate and overall number of Toffoli gates. This effect is exacerbated by their larger ancilla count---even for adders with equivalent Toffoli counts and depths, additional ancilla qubits can be viewed as incurring an opportunity cost measured in units of Toffoli gates, as the spacetime volume they occupy could otherwise be used in MSFs \cite{gidney2020QuantumBlockLookahead}. Ultimately, for relevant values of $n$ in applications, the superior constant prefactors on the gate complexity of ripple-carry addition lead it to remain the default choice in most compilations.
An exception to this is the spacetime-optimized resource estimates of Shor's algorithm in Refs.~\cite{gidney2021HowToFactor,zhou2025resourceAnalysisLowOverheadTransversal}, which performed in-place addition on integers with $n=2048$ bits using a ripple-carry adder with $r=1$ \cite{gidney2021HowToFactor} and $r=20$ \cite{zhou2025resourceAnalysisLowOverheadTransversal}  oblivious carry runways, reducing the depth roughly by a factor of up to $r+1$ in each case. Additionally, those compilations implemented the ripple-carry addition in a time-optimal, reaction-limited fashion; it is well suited for this method because of the sequential nature in which it computes the carry bits. Reaction-limited quantum addition enables a reduction in the time for each addition step without moving to a fundamentally lower-depth adder. 

It is worth emphasizing that regardless of which quantum adder compilation is chosen, the logical gates in the compilation are then implemented in a general-purpose way as prescribed by the particular paradigm for FTQC, such as those discussed in the previous section. For example, the space-optimized compilation of Shor's algorithm in Ref.~\cite{gidney2025factor} performs ripple-carry addition on integers of size up to $n=33$ by writing the circuit as a sequence of $2n-2$ lattice surgery steps (meanwhile consuming $n-1$ CCZ magic states). At $d=25$ and assuming \SI{1}{\micro \second} per SE round, the total time for the addition to be completed is estimated to be \SI{1.6}{\milli \second}.\footnote{By switching to a carry-lookahead adder with reaction depth roughly $4 \lg(n) \approx 20$, and running the adder at the reaction limit (assuming a reaction time $\tau_r = \SI{1}{\micro \second}$), the total time for the 33-bit quantum addition could be reduced to a minimum of roughly \SI{20}{\micro \second}, at the expense of significantly larger footprint for logical ancilla qubits (required for carry-lookahead addition), additional magic state factories to match higher Toffoli consumption rate, and teleportation modules for time-optimal computation (see App.~\ref{App:time_optimal_computation}). } In contrast, in classical hardware, specialized adder circuits (typically using logarithmic-depth approaches) are hard-coded into the arithmetic logic unit; signals propagate through this circuit on each clock cycle of the CPU, allowing for natural parallelization of the classical gates composing the adder (in the sense that all are performed within the same clock cycle), at least up to fixed input size, such as $n=64$. The total time for the addition to complete is thus less than \SI{1}{\nano \second}, more than one million times faster than the estimate for ripple-carry quantum addition implemented via lattice surgery.

\subsection{Quantum lookup tables and quantum read-only memory}\label{Subsec:LookupTablesQROM}

A quantum lookup table is an operation for loading  an entry of a classical dataset $D$ into a quantum state in superposition. Let $D_x$ be the $b$-bit data item located at address $x \in \{0,1,\ldots, N-1\}$ of the data table. The goal is to implement the unitary operation that, for an arbitrary set of complex coefficients $\{\alpha_x\}$, maps
\begin{align}\label{eq:quantum_lookup_table}
    \sum_{x=0}^{N-1} \alpha_x \ket{x}\ket{0} \mapsto \sum_{x=0}^{N-1} \alpha_x \ket{x} \ket{D_x}\,.
\end{align}
This operation is used in a number of quantum algorithms. For example, quantum arithmetic can be made cheaper by classically pre-computing the arithmetic over a specified range of inputs, and then coherently retrieving the results from a lookup table as needed. This technique, known as windowed quantum arithmetic, has been applied to reduce the resources needed for Shor's algorithm \cite{gidney2019windowedQuantumArithmetic,gidney2021HowToFactor,litinski2023EllipticCurvesBaseline}. In quantum simulation applications, quantum lookup tables are used to read in coefficients of Hamiltonians for molecules or other physical systems. In these contexts, the proposed implementation of the lookup table of Eq.~\eqref{eq:quantum_lookup_table} is typically a method referred to as quantum read-only memory (QROM) \cite{babbush2018EncodingElectronicSpectraLinearT}, where the $N$ entries of the table are queried sequentially. The cost of this approach is only $\lg(N)$ ancilla qubits and $N-1$ (serial) Toffoli gates (or $4N-4$ $T$ gates). At the expense of 25\% more Toffoli gates, the $\lg(N)$ ancilla qubits can be dirty,\footnote{A dirty qubit is a logical qubit with an unknown initial state (possibly entangled), and can be used as an ancilla qubit for computation, provided it is returned to its (unknown) initial state.} or with 125\% more Toffoli gates, the number of ancilla qubits can be reduced to essentially $\mathcal{O}(1)$ \cite{khattar2024riseConditionallyCleanAncillae}.

On the other hand, by adding more ancilla qubits (most of which can be dirty), the non-Clifford gate count of implementing the quantum lookup table can be reduced, a strategy that has been referred to by the acronym QROAM \cite{Berry2019QubitizationOfArbitraryBasisChemistry}. Specifically, in Ref.~\cite{low2018tradingTgatesforDirtyQubits}, a method is given that, for any choice $\lambda\leq N$, implements the quantum lookup table with $8\lceil N/\lambda\rceil +32 b \lambda$ $T$ gates, $b$ clean ancilla qubits, and $b \lambda$ dirty ancilla qubits. The minimal $T$ count of $\Theta(\sqrt{Nb})$ is achieved at $\lambda = \Theta(\sqrt{N/b})$. In paradigms where non-Clifford gates dominate the cost of the computation and are implemented sequentially, this represents the fastest way of implementing the quantum lookup table. Indeed, the advent of QROAM has been one of several significant factors responsible for the dramatic reduction in resource estimates for quantum chemistry. For example, Ref.~\cite{Berry2019QubitizationOfArbitraryBasisChemistry} showed how using QROAM together with a large number of ancillas could result in a $20 \times$ reduction in the Toffoli count, compared to a smaller-footprint approach with limited usage of the QROAM idea. It is worth noting that QROAM has some natural parallelism---the $T$ depth is only $\mathcal{O}(N/\lambda + \log(\lambda))$, which can be much smaller than the $T$ count at large values of $\lambda$, an observation exploited in Ref.~\cite{kim2022FaultTolerantQuantumChemicalSimulationsLiIon} to reduce the wall time of the quantum lookup table for simulations of battery molecules.  Furthermore, through alternative techniques, a reduction in $T$ count to $\mathcal{O}(\sqrt{N})$ can  be realized together with a reduction in total circuit depth  to $\mathrm{polylog}(Nb)$ \cite[Appendix C]{low2018tradingTgatesforDirtyQubits} at the expense of $\Theta(Nb)$ total (dirty) ancilla qubits. If one only cares about the non-Clifford depth, then $\mathcal{O}(\log\log(N))$  $T$ depth is possible, and this can even be achieved together with $\mathcal{O}(\sqrt{Nb})$ $T$ count and $\mathcal{O}(\sqrt{Nb})$ ancilla count \cite{mukhopadhyay2024qRAMloglogTdepth}, although in that case requiring at least $\Omega(\sqrt{Nb})$ Clifford depth.

Even when the number of non-Clifford gates is $\Theta(\sqrt{N})$, the total number of (Clifford plus non-Clifford) gates must always be at least $\Omega(Nb)$; this fact is a consequence of the observation that the dataset $D$ has $Nb$ degrees of freedom. It is possible to arrange these $\Theta(Nb)$ gates into a circuit with only $\Theta(\log(Nb))$ total circuit depth by using $\mathcal{O}(Nb)$ ancilla qubits, and if structured correctly these circuits may even demonstrate some intrinsic noise resilience \cite{hann2021resilienceofQRAM}. When implemented in logarithmic depth, the operation of Eq.~\eqref{eq:quantum_lookup_table} is often referred to as quantum random access memory (QRAM) \cite{giovannetti2007QuantumRAM}---it may be viewed as a quantum analogue of classical RAM, where  an entry of a large classical database can be accessed with latency that grows only very weakly with increasing memory size. Indeed, implementing Eq.~\eqref{eq:quantum_lookup_table} with latency $\mathrm{polylog}(Nb)$ is necessary to realize quantum advantage in certain machine learning applications, such as recommendation systems \cite{kerenidis2016QRecSys} and Gaussian process regression \cite{zhao2019TrainingGaussianProcess}, and generally in any big-data application where purported quantum speedups come from the ability of quantum algorithms to explore large datasets in superposition. However, implementing these shallow-but-wide QRAM circuits within the FT paradigms discussed here would require an overwhelmingly large device footprint (see, e.g., resource estimates in \cite{dimatteo2020FaultTolerantQRAM}) and a burdensome amount of parallel quantum error correction, leading to criticism that applications relying on cheap QRAM will never be practical \cite{aaronson2015ReadTheFinePrint, jaques2023qram}. In response to this objection, alternative paradigms specialized for implementing low-latency large-scale fault-tolerant QRAM have been suggested \cite{dalzell2025distillationteleportationprotocolfaulttolerantqram}, distinct from the general-purpose FTQC paradigms considered in this paper.

Regardless of the implementation, logical circuits for data lookup require long-range connectivity between logical qubits. Several works have investigated 2D layouts of data lookup circuits that would minimize routing issues and enable their implementation within the FTQC paradigms considered here \cite{haner2022SpacetimeLookupTable, xu2023systemArchitectureQRAM,zhu2024UnifiedLookup}. Other work \cite{cesa2025fastErrorCorrectableQuantumRAM} has proposed a method which redistributes much of the routing complexity of low-latency data lookup to offline preparation of an entangled $\mathcal{O}(N)$-qubit resource state. The resource state can be consumed at query time using Bell basis measurements and adaptive single-qubit Pauli measurements, resembling an application of the philosophy of time-optimal computation to QRAM. Optimizing the cost of table lookups requires balancing magic state consumption (distillation resources and routing) and data writing (Clifford) operations~\cite{lee2021EvenMoreEfficientChemistryTensorHyp,caesura2025ActiveVolumeChemistry}.

\subsection{Parallelization at the quantum algorithm level}

The spacetime tradeoffs discussed so far constitute optimizations performed at the `quantum circuit' and `logical circuit' levels of the stack depicted in Fig.~\ref{fig:AbstractCompilation}. 
We may also consider optimizations at the `quantum algorithm' layer. Quantum algorithms typically obtain speedups by making asymptotically fewer calls to certain black-box subroutines, or `oracles', than an analogous classical algorithm would need to make. Subroutines, such as adders and lookup tables, then appear in the implementation of these oracles. 
In seeking to minimize the wall time of quantum applications, we may wish to parallelize not only the subroutines implementing each black-box query, but also the queries themselves. In some instances, this is possible, while in others, lower bounds on query depth establish hard limitations.  

One example of a primitive which has been studied in this capacity is Hamiltonian simulation---that is, implementing the time evolution unitary $e^{iHt}$ given black-box access to a Hamiltonian $H$. It has long been known that the minimum query complexity of Hamiltonian simulation  grows linearly in $t$ \cite{berry2005EffQAlgSimmSparseHam}, which corresponds to the intuitive statement that time evolution cannot generally be fastforwarded (except in special cases where $H$ has certain exploitable structure \cite{atia2017fastForwarding,Gu2021fastforwarding}). It turns out that this asymptotic lower bound holds not only for the total number of queries to $H$, but also for the depth of those queries: it is not in general possible to reduce the query depth of Hamiltonian simulation simply by using more qubits, since $\Omega(t)$ queries must be done serially \cite{chia2023ImpossibilityParallelFastForwarding}. 
A similar query-depth lower bound asymptotically matching the total query count upper bound has also been derived for the quantum linear system problem in the black-box setting \cite{wang2024depthLowerBoundLinearSystems}. 
Complementary to this, it was shown that Hamiltonian simulation \textit{can} be parallelized with respect to increasing precision---while $\Omega(\log(1/\epsilon)/\log\log(1/\epsilon))$ queries are needed to achieve error $\epsilon$ \cite{berry2013ExpPrecHamSimSTOC}, these can be parallelized to exponentially lower $\mathrm{polylog} \log(1/\epsilon)$ depth \cite{zhang2024parallelHamiltonianSimulation}. 

Another example is Grover’s algorithm \cite{grover1996QSearch} and its generalizations.  Grover’s algorithm finds a marked item among $N$ items using $\mathcal{O}(\sqrt{N})$ queries to a black box that coherently computes whether an item is marked, achieving a quadratic speedup over classical exhaustive search, which requires $\Omega(N)$ queries. However, classical exhaustive search is embarrassingly parallel, and can be performed in $\mathcal{O}(\log(N))$ depth given access to $\mathcal{O}(N)$ parallel processors---each processor queries a different item, and using a tree-like communication network, they may funnel the index of the correct item to the desired output location. In contrast, the quadratic speedup of Grover’s algorithm can only be obtained on the portion of the algorithm performed serially \cite{zalka1999grover}: given a factor $k$ of additional space, the best one can do is split the $N$ items into $k$ groups of size $N/k$, and run $k$ copies of Grover’s algorithm independently in parallel on each group, achieving query depth $\mathcal{O}(\sqrt{N/k})$. As $k$ becomes larger and more parallel resources become available, the parallel quantum and classical algorithms converge in both time and space complexity. This illustrates a fundamental tension between some prominent sources of quantum speedup and parallelizability.

On the other hand, additional space can sometimes lead to reductions not only in the query depth but also in the query count. For example, consider the task of estimating to small additive error $\epsilon$ the expectation value of $M$ (generally non-commuting) observables in a certain fixed $n$-qubit state $\ket{\psi}$, given an oracle that prepares $\ket{\psi}$. For example, $\ket{\psi}$ may represent the ground or thermal state of a many-body quantum system. A minimal-footprint approach using $n+\mathcal{O}(1)$ qubits achieves $\mathcal{O}(M/\epsilon)$ queries by running amplitude estimation \cite{knill2007ObservableMeasurement} for each of the $M$ observables in series.  With $\mathcal{O}(M\log(1/\epsilon))$ additional qubits, one can learn all $M$ expectation values in parallel with a quadratically reduced total query count of $\mathcal{O}(\sqrt{M}/\epsilon)$ \cite{huggins2022ExpectationValue}.

\section{Costing fault-tolerant applications}\label{Sec:CostingApplications}

\begin{figure}
    \centering
    \includegraphics[width=0.95\linewidth]{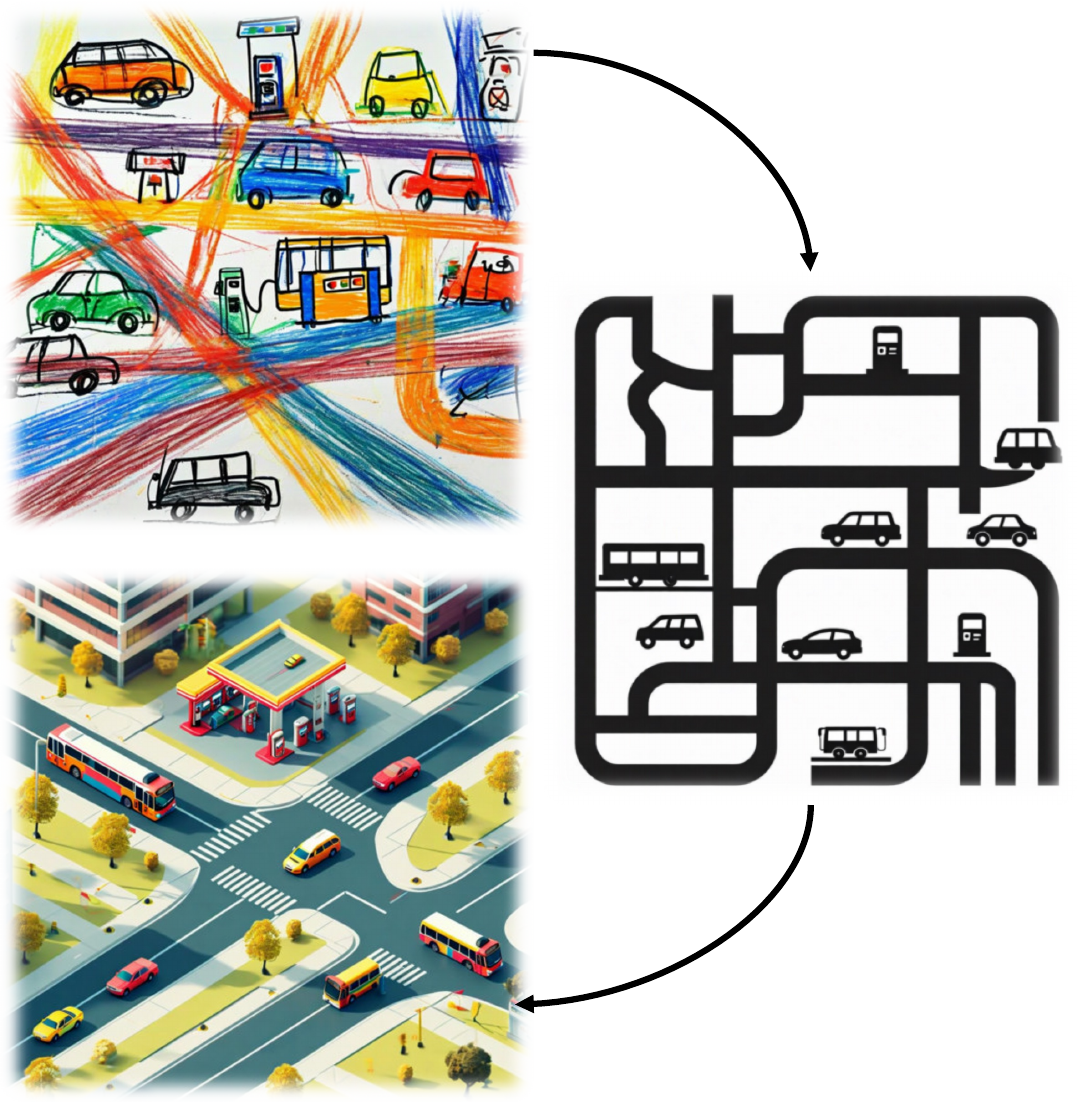}
    \caption{An analogy can be made between the complexity of compiling quantum algorithms and the complexity of planning a transportation network in a city. A city planner must maximize throughput of commuters using many choices of transport and routes, subject to constraints, such as cost of roads and placement of fuel stations. Simplifying the model and gradually adding in complexity can help to develop sensible heuristics and intuition for more efficient designs. Images generated using Amazon Nova Canvas.}
    \label{fig:TransportNetworks}
\end{figure}

In the previous sections we have outlined a number of spacetime tradeoffs available at both the algorithmic and FTQC levels. We have also discussed run time optimizations unlocked by additional hardware capabilities, for example, the availability of transversal logical entangling gates. However, the relative merits and compatibility of these different schemes are not obvious. Is it better to use a ripple-carry adder with AFT in a neutral atom platform, or to use a carry-lookahead adder with Clifford + $T$ compilation in a 2D superconducting processor? Such questions are best answered by quantitative numerical resource estimates with a specified algorithm, FTQC paradigm, and hardware model. However, as a precursor to this, we can consider simplified theoretical models to understand the combined effects of the different optimizations. These simplified models can make it easier to understand the sensitivity of resource estimates to different QEC and hardware parameters, as well as to the different choices of algorithmic subroutines and FTQC paradigms. In Fig.~\ref{fig:TransportNetworks} we draw an analogy between the challenges present in compiling quantum algorithms, and those encountered in other large-scale industrial design and optimization problems, such as planning a transportation network in a city.

\subsection{Simplified cost model for generic circuits}

As shown in Fig.~\ref{fig:AbstractCompilation}, it is challenging to simultaneously optimize algorithmic and FTQC choices. It is more natural to think in terms of stack layers, first optimizing algorithm design and then optimizing the compilation to FTQC operations. It is typically necessary to iterate the process to optimize resource estimates; an instructive example is the discussion of compilation in Ref.~\cite{lee2021EvenMoreEfficientChemistryTensorHyp}. In simplified models we can either assume a fixed algorithm and vary FTQC compilation, or assume a specified FTQC paradigm and optimize algorithm compilation. An example of the first case is Ref.~\cite{litinski2019gameofsurfacecodes}, which developed algorithm-agnostic models of running computations in the standard 2D PBC paradigm, clearly illustrating the available spacetime tradeoffs. An example of the second case is Ref.~\cite{litinski2022activeVolume}, which constructed models for the cost of common algorithmic subroutines, assuming computation via the active volume scheme. We will focus on the first case, considering a fixed algorithm and optimizing the underlying FTQC paradigms. 

\subsubsection{Clifford circuits}

A simple starting point is to consider a logical single- and two-qubit Clifford circuit with $N_c$ gates on $Q$ logical qubits. We assume gates are implemented in layers with an average parallelism factor $P_c$, such that there are $N_c/P_c$ layers. Note that $P_c$ might be lower than the parallelism present in the abstract logical circuit, because it is constrained by routing of operations. We can also consider implementing sequential gate layers in parallel using gate teleportation. Teleporting $M$ gate layers requires roughly $2 (M-1)\cdot P_c$ additional logical qubits, and reduces the run time by a factor of $M$, as discussed in App.~\ref{App:time_optimal_computation}. Our simplified model ignores the time cost of logical Bell state preparation and measurement. If this cost is included, its effect is to make teleportation unfavorable for small $M$ (e.g., $M<3$ if all operations are implemented via lattice surgery). Parallelizing gates (directly or via teleportation) introduces a $P_c,M$-dependent routing space cost $R(P_c,M)$. The routing cost function $R$ will also depend on other factors, such as the locality of logical gates compared to the layout of logical qubits. The overall physical space $\mathcal{S}$ and time $\mathcal{T}$ resources are given by:
\begin{align}
    \mathcal{S} &= q(d) \left[Q + R(P_c,M) + 2(M-1)P_c\right] \\
    \mathcal{T} &= \frac{N_c}{M P_c} \tau_c(d) 
\end{align}
where $d$ is the code distance, $\tau_c$ is the time for a logical Clifford gate and $q$ is the number of physical qubits per logical qubit. The code distance depends polylogarithmically on the logical spacetime volume, the target logical failure rate $\epsilon$ and the physical error rate $p$. By examining the spacetime volume $\mathcal{V} = \mathcal{S} \cdot \mathcal{T}$, we can make the following observations:
\begin{itemize}[itemsep=-1pt, topsep=6pt]
    \item Varying the amount of gate teleportation ($M$) constitutes a spacetime tradeoff in the regime where the resources for gate teleportation dominate the spacetime volume. In that regime, the total volume depends only on the number of Clifford gates $N_c$. 
    \item In the regime where we consider a modest amount of gate teleportation such that $MP_c \ll Q$, gate teleportation reduces $\mathcal{T}$ by a factor $M$, while only increasing $\mathcal{S}$ incrementally. Thus it reduces the overall spacetime volume.
    \item Platforms with the ability to reshuffle logical qubits will likely incur a smaller (possibly zero) routing overhead $R$, compared to platforms with planar logical connectivity. However, the time for a Clifford gate $\tau_c$ must account for the shuttling time, making this another form of spacetime tradeoff.
    \item AFT, which reduces $\tau_c$ by a factor of $\mathcal{O}(d)$ for 2D topological codes, improves space and time costs, and is compatible with logical gate teleportation. In contrast, moving to a 3D single-shot code reduces $\tau_c$ by a factor of $\mathcal{O}(d)$, at the cost of increasing $q$ by a factor of $\mathcal{O}(d)$.
\end{itemize}

\subsubsection{General circuits}
The model becomes more complex when considering non-Clifford gates. This is because of the need to prepare and consume magic states. We consider the same model for Clifford circuits described above, and augment it with $N_{nc}$ non-Clifford gates on $Q$ logical qubits. We assume that Clifford and non-Clifford gates are implemented in separate layers. Non-Clifford gates are implemented in layers with an average parallelism factor $P_{nc}$, such that there are $N_{nc}/P_{nc}$ layers. The time for each non-Clifford layer is given by $\tau_{nc}$. For example, teleporting $T$ states has cost roughly $\tau_{nc} = 2\tau_c + \tau_r$, where the factor of two accounts for the corrective $S$ operation. In contrast, operating at the reaction limit sets $\tau_{nc} = \tau_r$. We can implement a non-Clifford layer if we have access to $P_{nc}$ magic states. We define the average time to prepare $P_{nc}$ magic states as $\tau_m$. The number of physical qubits used for magic state preparation is $\frac{\tau_f}{\tau_m}q_fP_{nc}$ where $\tau_f$ is the time for a single $q_f$-qubit factory to prepare a magic state. If the time to distill a layer of $P_{nc}$ magic states is greater than the time interval between non-Clifford layers ($\tau_m > \frac{N_c \tau_c P_{nc}}{M P_c N_{nc}} + \tau_{nc}$) then the computation is bottlenecked by magic state production. The overall physical space $\mathcal{S}$ and time $\mathcal{T}$ resources are given by:
\begin{align}
    \mathcal{S} &= q \left[Q + R + \max[2(M-1)P_c ,k\frac{\tau_c}{\tau_r}P_{nc} ] \right] + \frac{\tau_f}{\tau_m}q_fP_{nc} \\
    \mathcal{T} &= \max\left[\frac{N_c \tau_c}{M P_c} + \frac{N_{nc} \tau_{nc}}{P_{nc}}, \frac{N_{nc}\tau_m}{P_{nc}}\right].
\end{align}
The factor $k\frac{\tau_c}{\tau_r}P_{nc}$ accounts for additional space resources for reaction-limited computation, and so $k$ is either a small constant if reaction-limited computation is used, or zero otherwise. The physical factors $q, q_f$ and $\tau_c, \tau_f, \tau_r$ all depend on the physical error rate $p$ and the target logical failure rate $\epsilon$. We can make the following observations:
\begin{itemize}[itemsep=-1pt, topsep=6pt]
\item Many existing resource estimates proceed as if $q_f \tau_f$ is large compared to $q \tau_{nc}$. To maintain small physical footprints, they use a single factory such that $\tau_m=P_{nc}\tau_f$. The calculation is likely bottlenecked by magic state preparation, favoring a sequential compilation with $P_{nc}=1$, which typically minimizes $N_{nc}$. The space cost is typically dominated by logical memory $qQ$.
\item Assume no additional teleportation of Clifford gates (i.e., $M=1$), and that the circuit is not bottlenecked by magic state production. We can define the ratio of the Clifford depth to the non-Clifford depth as $C^* := \left( \frac{N_c \tau_c}{P_c} \right)/\left(\frac{N_{nc} \tau_{nc}}{P_{nc}} \right)$. The value of $C^*$ will depend on the chosen FTQC paradigm. PBC is recovered by compiling away Clifford gates such that $N_c=0$. As discussed previously, this typically reduces the parallelism of non-Clifford gates such that $P^{PBC}_{nc}=1$. Under these assumptions, we can consider the ratios of the run times of circuits with and without PBC. Assume that $M=1,k=0$. The ratio $\mathcal{T}^{PBC}/\mathcal{T} = P_{nc}/(1+C^*)$. As may be expected, PBC increases the run time if the reduction in Clifford depth is outweighed by the loss of parallelism in the non-Clifford layers. 
\item For PBC, setting $\tau_m = \tau_{nc}$, $P_{nc} = 1$, $k=0$ and $R = (Q + \sqrt{8Q} +1)$ recovers the fast block PBC architecture of Ref.~\cite{litinski2019gameofsurfacecodes} which assumes that non-Clifford gates are implemented sequentially and the run time depends only on the number of non-Clifford gates.
\item In light of recent work (e.g.,~\cite{gidney2024magiccultivation}) to reduce $\tau_f q_f$, it is possible to set $\tau_m=\tau_{nc}$ and to increase $P_{nc}$ without substantially increasing total qubit counts. The spacetime volume for magic state preparation is $N_{nc} \tau_f q_f$, and so only depends on the total number of non-Clifford gates and the characteristics of the factory. In particular, this means the spacetime volume spent on magic state preparation is independent of whether non-Clifford gates are implemented in parallel, or if reaction-limited computation is used (though note that the space allocated to magic state preparation is dependent on these choices). Assuming that the calculation is not magic state bottlenecked, the ratio of magic state preparation volume to logical storage volume is $\left(\frac{P_{nc}}{(1+C^*)}\right) \left(\frac{\tau_f}{\tau_{nc}}\right)\left(\frac{q_f}{qQ}\right)$ (when we set $M=1,k=0$). 
\end{itemize}

In reality, the alternating layers of Clifford and non-Clifford gates will have varying depths (unless compiled to PBC). This introduces complexity, such as variably allocating routing space and MSF space. Furthermore, in real algorithms the values of $Q$, $P_c,P_{nc},N_c, N_{nc}$ are related through the choice of algorithmic subroutines, e.g., QROM vs.~QROAM, ripple-carry vs.~carry-lookahead adders. These choices are best captured in a simple model through subroutines with a smooth tradeoff in resources, e.g., block-lookahead adders. Similarly, different algorithmic subroutines may have different levels of logical locality, which impacts the values of $P_c, P_{nc}, R$. The large search space to optimize over makes this a challenging problem for human scientists, and motivates the development of automated compilation programs tailored for the underlying hardware platforms.

\subsection{Example: Simulating the Fermi-Hubbard model}\label{Sec:FH_results}

In this section, we present resource estimates for simulating time evolution of the 2D Fermi-Hubbard (FH) model. These resource estimates quantify the spacetime costs of parallelizing a calculation, illustrating a selection of the ideas presented in previous sections. The FH Hamiltonian describes fermions hopping on an $L \times L$ 2D lattice
\begin{equation}
    H = -t\sum_{\langle i,j\rangle, \sigma}^L \left(a_{i, \sigma}^\dagger a_{j, \sigma} + a_{j, \sigma}^\dagger a_{i, \sigma} \right) + \frac{U}{4}\sum_{i}^L Z_{i, \uparrow} Z_{i, \downarrow}
\end{equation}
where $\langle i,j\rangle$ enumerates edges on the 2D lattice, $a_{i,\sigma}^\dagger$ and $a_{i,\sigma}$ are creation and annihilation operators for fermions at lattice site $i$ with spin $\sigma \in \{\uparrow,\downarrow\}$, and  $Z_{i,\sigma} = (I-2a_{i,\sigma}^\dagger a_{i,\sigma})$ results from a shift of chemical potential.\footnote{The Hamiltonian conserves the number of fermions of each spin, so for initial states with known values of spin up/down fermions shifting the chemical potential contributes an unimportant global phase to the time evolution~\cite{CampbellHubbard22}.} Tunable parameters $t$ and $U$ denote the strength of the hopping term and the strength of the on-site interaction, respectively. The FH model is studied as a simplified model of materials, with applications in high-temperature superconductivity~\cite{hubbard1963Hubbard,leblanc2015TwoDimHubbard,fradkin2015ColloquiumHighTcSC}, spintronics~\cite{zutic2004Spintronics}, photonic control~\cite{oka2019FermiHubbardDynamics}, and studies of thermalization~\cite{polkovnikov2011NonEquilibriumDynamics}, and it has become a popular target and benchmark for quantum simulation algorithms~\cite{wecker2015StronglyCorrelated,babbush2018EncodingElectronicSpectraLinearT,Kivlichan2020ImprovedFaultTolerantSimulationCondensedMatter,CampbellHubbard22,flannigan2022,yoshioka2022CondensedMatterSimulation}. Guided by the workflow in Fig.~\ref{fig:AbstractCompilation}, we investigate a number of compilation choices. At the algorithmic level of the stack, we examine two quantum algorithms for simulating time evolution: plaquette Trotterization (PLAQ)~\cite{CampbellHubbard22} and quantum signal processing (QSP)~\cite{low2016HamSimQSignProc,low2016HamSimQubitization}. PLAQ is a naturally parallelizable approach which divides the Hamiltonian evolution into simulation of local commuting plaquette terms. For PLAQ, we consider three varying compilations at the quantum and logical circuit levels, which consume $T$-gates at different rates: 1) Serially, 2) $L$ $T$-gates in parallel, 3) $L^2$ $T$-gates in parallel. These are compared against a parallel compilation of QSP that uses a logarithmic depth circuit for block-encoding the FH Hamiltonian~\cite{Wan2021exponentiallyfaster}. QSP is a post-Trotter method that scales asymptotically optimally with the evolution time $\T$ and simulation error $\epsilon$. We explain the algorithms and compilations in detail for PLAQ and QSP in App.~\ref{App:PlaqTrot}~\&~\ref{App:QSP}, respectively. Our results illustrate how achieving parallelism at the logical circuit level requires co-design with algorithmic choices, such as the fermion-to-qubit mapping.

\begin{figure}
    \centering
    \includegraphics[width=0.95\linewidth]{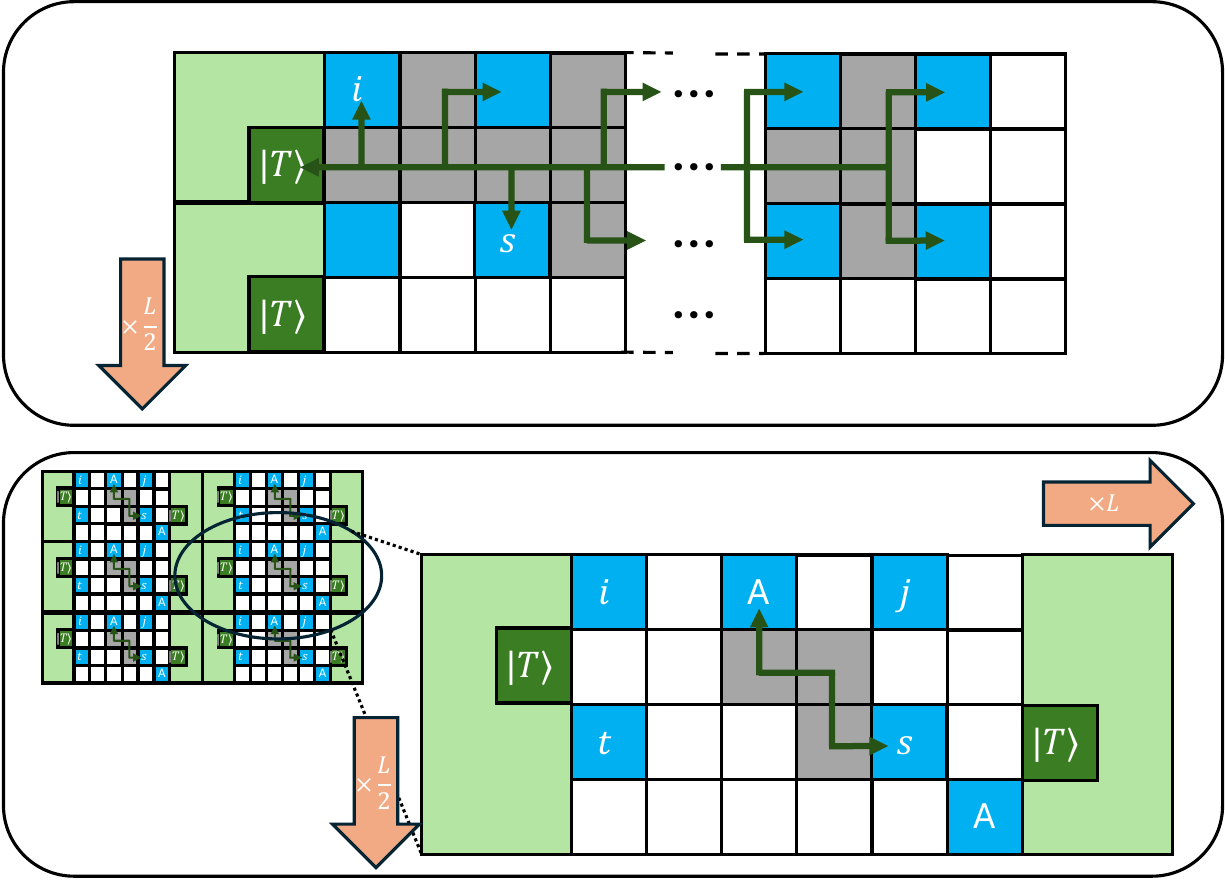}
    \caption{Surface code layouts for `$L$-parallel' (upper) and `$L^2$-parallel' (lower) compilations of PLAQ. Logical data qubits representing fermion lattice sites are in blue, magic state factories (MSFs) in light green, magic states in dark green, and idling/active routing space in white/grey. The layout for $L$-parallel shows one row of plaquettes of a given spin, and should be repeated vertically as indicated by the orange arrow (and mirrored for spin). MSFs are located at the boundary, saturating the rate of magic state consumption. The diagram shows a multiqubit operation used to implement the hopping term. The layout for $L^2$-parallel shows a single unit cell (magnified), and should be repeated both horizontally and vertically (indicated by the orange arrow), embedding MSFs into the lattice (inset). The additional logical data qubits $A$ are the auxiliary qubits required for the local fermion-to-qubit mapping. The diagram shows part of a swap operation between fermions $j$ and $s$, mediated by the auxiliary site.}
    \label{fig:FH_layouts}
\end{figure}

We compile to the standard 2D architecture discussed in Sec.~\ref{Subsec:Standard2D}, which encodes logical qubits into the 2D surface code, and implements logical operations via lattice surgery. We obtain estimates for the logical spacetime volume of the circuits by compiling to the layouts shown in Fig.~\ref{fig:FH_layouts}, optimizing the scheduling of lattice surgery operations, as well as the rate of magic state distillation and consumption.

As stated above, we consider three compilations of PLAQ, motivated as follows.
\begin{enumerate}[itemsep=-1pt, topsep=6pt]
    \item Serial compilation proceeds via PBC to eliminate Clifford gates, and uses Hamming weight phasing (HWP)~\cite{gidney2018_halving_addition,Kivlichan2020ImprovedFaultTolerantSimulationCondensedMatter} to reduce $T$ count. The use of HWP with PBC increases the support of each $\pi/8$ PPR, reducing available parallelism. A minimal number of MSFs are required. Each Trotter step has depth $\mathcal{O}(L^2)$.
    \item `$L$-parallel' uses the Jordan-Wigner (JW) fermion-to-qubit mapping and explicitly implements Clifford gates via lattice surgery. The JW mapping on $N$ fermionic modes encodes local fermionic operators as qubit operators with support on $\mathcal{O}(N)$ qubits, and so is considered nonlocal. The growth of locality under the JW mapping can reduce the ability to parallelize the underlying circuit, as operations that were disjoint in fermionic space may now overlap in qubit space. In our compilation, the rate of magic state consumption is chosen to match the locality of Jordan-Wigner strings, as well as the rate of magic state teleportation from $\mathcal{O}(L)$ MSFs placed at the boundary of the computational region~\cite{beverland2022SurfaceCodeCompilation}. Each Trotter step has depth $\mathcal{O}(L\log(L))$, where the additional logarithmic overhead arises due to synthesizing rotations from $T$ gates.
    \item `$L^2$-parallel' uses a local fermion-to-qubit mapping with $1.5$ qubits per fermionic mode~\cite{derby2021CompactMapping} to eliminate long-range JW strings that otherwise limit parallelism. MSFs are co-located with data qubits to enable parallel consumption of $L^2$ $T$-gates. Each Trotter step has a depth of $\mathcal{O}(\log(L))$, which stems from the synthesis of rotations from $T$ gates.  
\end{enumerate}

We use a co-design mentality to increase the degree of parallelization in our compilations. By examining constraints at the algorithmic, circuit, and architectural levels, we ensure that parallelization is not bottlenecked by a single factor.

\subsubsection{Comparing compilations}
\begin{figure}[!h]
\begin{center}
\includegraphics[width=\columnwidth]{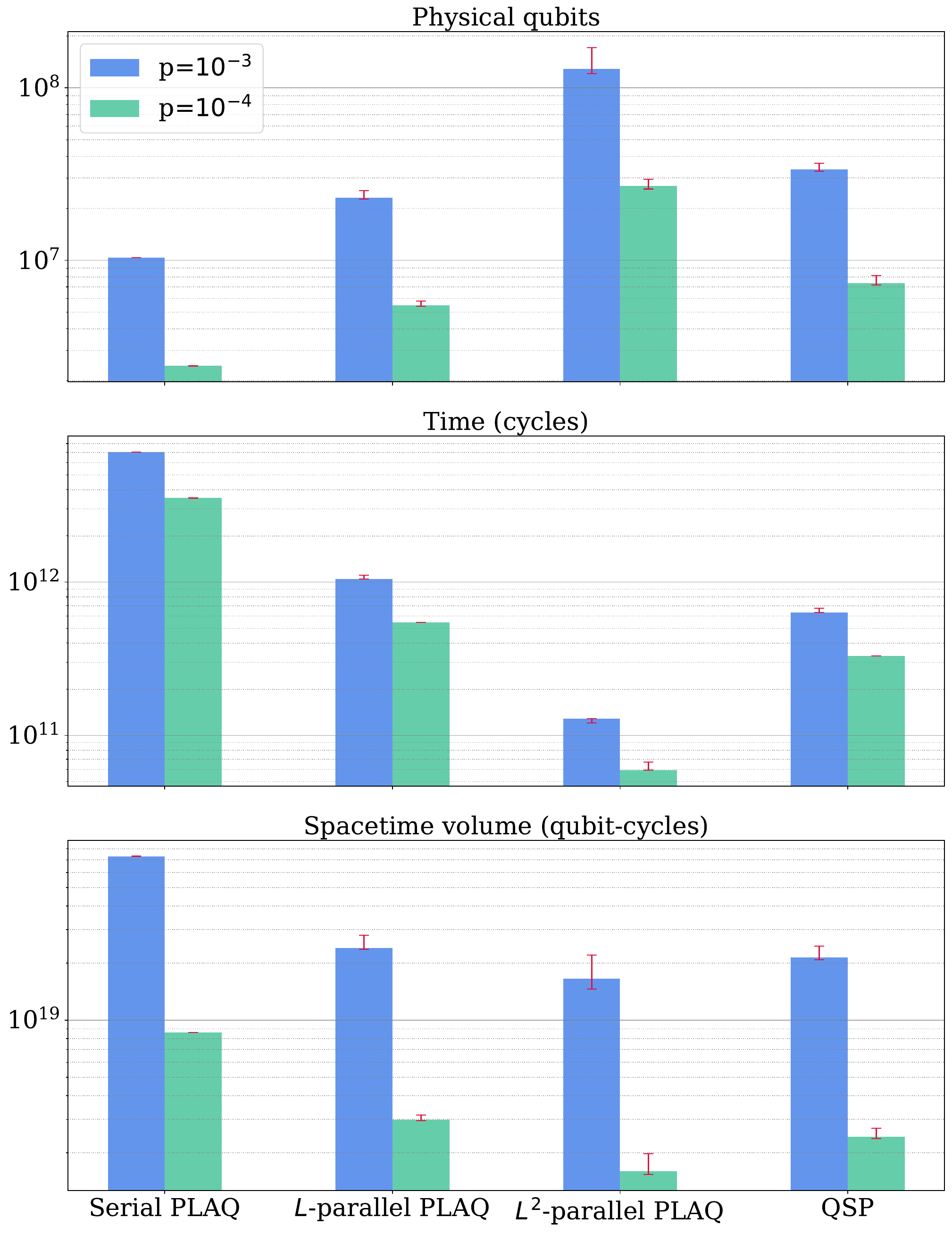}
\caption{Resource estimates for the 2D FH model with parameters $L=30, \T=300, U/t=8, \epsilon=0.01$, at physical error rate $p$. Error bars show resources assuming a $\pm5$\% change in time and qubit requirements to distill a magic state, and in the surface code threshold and logical error rate prefactor.
}
\label{Fig:ResourceEst}
\end{center}
\end{figure}

An example resource estimate is shown in Fig.~\ref{Fig:ResourceEst} for a 2D FH lattice with $L=30$ and evolution time $\T=10L$. Note that this is a larger system size and longer time duration than typically considered in literature resource estimates~\cite{babbush2018EncodingElectronicSpectraLinearT,Kivlichan2020ImprovedFaultTolerantSimulationCondensedMatter,yoshioka2022CondensedMatterSimulation,flannigan2022}, and would likely be well beyond the reach of classical computation. However, this leads to large resource estimates with approximately $10^{11}$--$10^{12}$ $T$ gates. In order to achieve the fastest run times and minimal spacetime volumes, it is necessary to move away from PBC and implement Clifford gates explicitly. The ability to execute gates in parallel eliminates idle spacetime volume, thus reducing the total spacetime volume---as well as reducing run time. This comes at a cost of a larger physical footprint, resulting from additional MSFs, routing space, and auxiliary qubits.

It is interesting to compare the $L^2$-parallel Trotter and QSP results. It is conventional wisdom in the field that QSP methods are faster, but require more qubits than Trotter approaches. Our results refute this for the most parallelized calculations. Asymptotically, 2nd order Trotter scales as $\mathcal{O}(L^{2.5}\mathcal{D})$ for our simulation parameters, where $\mathcal{D}$ is the depth of each Trotter step (see App.~\ref{App:TrotterError}). In contrast, QSP scales as $\mathcal{O}\left(L^3 \log(L) \right)$ (see App.~\ref{App:QSP}). For our most parallel PLAQ compilation, where $\mathcal{D} = \mathcal{O}\left(\log(L)\right)$, PLAQ has a lower asymptotic depth, which is already evident at $L=30$. This should be contrasted with an asymptotic scaling of $\mathcal{O}\left(L^{4.5} \right) $ for the serial compilation of PLAQ. Nevertheless, our $L^2$-parallel compilation of PLAQ requires 50\% more logical qubits to store the fermionic modes (due to the local fermion-to-qubit mapping), and requires additional MSFs and routing space to saturate the parallelism. This increases the physical qubit count beyond that of QSP. This highlights the importance of evaluating multiple algorithms and compilation schemes for a given task, and selecting the option that minimizes the chosen cost function, for the given architecture.

Our results are in good agreement with concurrent resource estimates that compared Clifford + $T$ and PBC compilations of Trotter and QSP for various spin systems~\cite{leblond2025ResourceComparison}. That work also observed reduced time and spacetime volume costs in Clifford + $T$--compiled Trotter circuits, compared to PBC-compiled Trotter. However, our resource estimates additionally parallelize QSP by using a less sequential SELECT oracle~\cite{Wan2021exponentiallyfaster}.

\subsubsection{Spacetime tradeoffs in applications}
We observe that the spacetime tradeoff of parallelization is better than linear; the factor by which run time is reduced is larger than the factor by which qubit count is increased. This can be compared to other available spacetime tradeoffs. For example, imagine we want to estimate a simple observable (e.g., site occupation) by averaging $\mathcal{O}(1/\epsilon^2)$ measurements in the computational basis. Given a sufficiently large processor, we could envisage running a number of small footprint (serial gates) calculations in parallel, thus requiring fewer repetitions of the calculation. Alternatively, we could run a single large-footprint (parallel gates) calculation. The first approach is only a linear spacetime tradeoff, and so is less efficient than the large-footprint, parallel gates approach. We note that other algorithmic strategies could be even more efficient, such as using amplitude estimation or coherently computing observables in parallel~\cite{huggins2022ExpectationValue,apeldoorn2022TomographyStatePreparationUnitaries}.

\subsubsection{Memory vs.~magic state factories}
Using typical models of performance for a surface code logical memory, lowering physical error rates from $10^{-3}$ to $10^{-4}$ leads to a $2 \times$ reduction in code distance $d$ and thus a $4\times$ saving in physical qubits~\cite{fowler2012SurfaceCodes,litinski2019gameofsurfacecodes}. However, the saving for magic state preparation can be much larger---as shown, for example, in Ref.~\cite[Table 1]{litinski2019magicstate}. As a larger fraction of the total footprint is allocated to MSFs in parallel compilations, we might expect a larger reduction in spatial resources when lowering physical error rates in parallel calculations than in serial calculations. There are several reasons this is not observed prominently in our results. First, using the results and notation from Ref.~\cite[Table 1]{litinski2019magicstate}, changing from the $(15:1)^6_{13,5,5} \times (15:1)_{29,11,13}$ MSF at $p=10^{-3}$ to the $(15:1)^4_{9,3,3} \times (20:4)_{15,7,9}$ MSF at $p=10^{-4}$ only provides a spacetime volume reduction of $\sim10\times$. This is a smaller reduction than some other $p=10^{-3}, 10^{-4}$ factory pairs. Second, we opt to consume magic states with lattice surgery in $\mathcal{O}(d)$ syndrome extraction rounds, which means that when the code distance $d$ is reduced by a factor of two, magic states are required twice as quickly. This implies only a $5\times$ saving in MSF space. Finally, the increased consumption of magic states in parallel calculations necessitates increased routing space, which we model in the same way as memory. This reduces the relative impact of MSFs to the total footprint reduction, compared to that of memory.

\subsubsection{Improvements to magic state preparation}
Magic state cultivation (MSC) has recently been introduced as a more efficient method of preparing $T$-states~\cite{gidney2024magiccultivation}. The logical error rates achieved by MSC are not yet sufficiently small for use in our resource estimates, which require approximately $10^{12}$ $T$-gates. Nevertheless, anticipating future improvements to MSC, we carried out resource estimates for the $L^2$-parallel simulations. In its current form, the success rate of MSC decreases as the fidelity of the output magic state increases. In order to maintain a constant rate of magic state consumption, we assume that many copies of MSC are run in parallel, to account for the probability of failure. Based on the values reported in \cite[Fig.~1]{gidney2024magiccultivation}, we assume a $25\times$ reduction in spacetime volume for MSC over MSD~\cite{litinski2019magicstate}, which we interpret as a $5\times$ reduction in qubit count and a $5\times$ reduction in time. At $p=10^{-4}$, we observe a $4.3\times$ reduction in total qubit count, due to the large number of MSFs required for that calculation. If improvements to MSC are not forthcoming, it would be interesting to consider the best ways to integrate MSC with existing distillation subroutines, in order to reach algorithmically relevant error rates.

\subsubsection{Reaction-limited computation}
The run time of the algorithm could be further reduced by incorporating reaction-limited quantum computation into our $L^2$-parallel compilation. However, we will show that the run time reductions come at the price of a large space overhead. As discussed in more detail in App.~\ref{App:time_optimal_computation}, reaction-limited computation implements a batch of $G_{\mathrm{opt}}$ sequential non-Clifford gates in time $\tau_r G_{\mathrm{opt}}$, using gate teleportation. The value of $G_{\mathrm{opt}}$ is chosen such that the reaction time delay associated with the gates is as much of a bottleneck as preparing the gate teleportation modules, i.e., $T_{\mathrm{Prep}} = G_{\mathrm{opt}} \tau_r$. There is no time overhead associated with teleporting in Clifford gates, but they do contribute to the space overhead of the protocol. 

The most straightforward subroutine to analyze is the synthesis of a rotation by applying a sequence of $ \approx 30$ $T$-gates. This is a similar circuit to the circuit for $HTHT$ shown in App.~\ref{App:time_optimal_computation}. We assume\footnote{Each teleportation module requires a Hadamard and CNOT gate, implemented via lattice surgery in $5d$ syndrome extraction rounds. We assume \SI{1}{\micro \second} per syndrome extraction round, and $d=15$.} each teleportation module can be prepared in $T_{\mathrm{Prep}} = $\SI{75}{\micro \second}, and a reaction time of $\tau_r =$ \SI{5}{\micro \second}. This implies $G_{\mathrm{opt}} = 15$, and so we will teleport the $T$-gates in two batches. The total time required is \SI{150}{\micro \second}. Reaction-limited computation uses 60 logical qubits ($G_{\mathrm{opt}} = 15$ teleportation modules, each with 4 logical qubits). For comparison, implementing the $T$-gates sequentially via lattice surgery takes \SI{825}{\micro \second} and uses two logical qubits\footnote{We assume $d$ rounds of syndrome extraction to consume a $T$-state using~\cite[Fig.7]{litinski2019gameofsurfacecodes}, $d$ rounds of syndrome extraction to implement an $S$ gate that is required half the time, and $\tau_r$ to determine if the $S$ gate is required. We require one logical qubit, plus one fresh magic state throughout the subroutine.}. Overall, we have a $5.5 \times$ reduction in time, at the cost of a $30\times$ increase in space. In this case, reaction-limited computation is a worse-than-linear spacetime tradeoff. The space overhead does not include any additional routing space, nor the additional MSFs required to distill $T$-gates at a faster rate.

\section{Conclusion and paths forward}

In this perspective, we have highlighted the need for faster logical clock speeds, and a number of strategies to achieve them. Although quantum algorithms are asymptotically more efficient than known classical algorithms for certain problems, quantum computers will be significantly slower than their classical counterparts.
This slowdown results from the slower operation speed of many types of physical qubits compared to transistors, the overhead of compiling classical floating-point operations to logical quantum operations, the overheads of QEC and FTQC, and the latencies involved in real-time decoding. The slowdown is exacerbated by current approaches to compilation. Compilation strategies which favor sequential implementation of gates, such as Pauli-based computation or using minimal magic state factories, lead to idle spacetime volume and unnecessarily slow run times. Parallelized compilations are necessary to eliminate this idle spacetime volume, as highlighted in our resource estimates for simulating the Fermi-Hubbard model, as well as by concurrent work performing similar resource estimates for spin systems~\cite{leblond2025ResourceComparison}.

Optimizing logical clock speeds may have a number of implications for the way we envisage building quantum computers and running quantum algorithms. First, in the long term we advocate for transitioning away from Pauli-based computation, in order to implement gates in parallel. This results in the need to implement Clifford gates explicitly. This is supported by recent cost reductions for preparing magic states~\cite{gidney2024magiccultivation}, which demonstrate comparable cost to logical Clifford gates. In this paradigm, more attention must be paid to the problem of scheduling and routing logical operations. Second, we may double-down on codes with access to fast transversal gates, such as surface codes and color codes, despite their suboptimal encoding rates. Third, implementation of real-time QEC with sufficiently high throughput and latency may favor using simple decoders (e.g., cellular automata), which may not be compatible with the most resource-efficient QEC codes.

Future resource estimates should be realistic about the logical clock speeds required to achieve quantum utility in industrial and scientific workflows---and the implications for physical qubit counts. Recent work has focused on reducing physical qubit counts by exploiting noise bias or higher-rate QEC codes. Ultimately, the saved space may need to be reinvested into spacetime tradeoffs which boost the logical clock speed, such as parallel production of magic states, more routing space, or reaction-limited computation. Even for the fastest qubit modalities, such as superconducting qubits, it will be challenging to achieve the required logical clock speeds. This challenge is even more daunting in slower modalities, such as neutral atoms or trapped ions. Further work is needed to improve logical clock speeds in those platforms, in the spirit of recent work on algorithmic fault-tolerance~\cite{Zhou2024AlgorithmicFT,cain2025FastCorrelatedDecoding}. For now, the best quantum computer is the one you can actually build. But in the long term, the best quantum computer may be the one that runs the fastest.

\section*{Acknowledgements}
We thank Arne Grimsmo and Joe Iverson for helpful comments on this manuscript. We thank Craig Gidney for answering questions on reaction limited computation, and Earl Campbell for answering questions on plaquette Trotterization.

\bibliography{Bibliography,bib_AK}

\appendix

\section{Parallelism through gate teleportation and time-optimal computation}\label{App:time_optimal_computation}

Gate teleportation can be used to reduce run times of quantum circuits, at the cost of increased spatial resources. Gate teleportation is a probabilistic process, potentially requiring a corrective operation. Assuming the teleported gate is in the Clifford hierarchy, the corrective operations are one level lower in the Clifford hierarchy than the teleported operation \cite{gottesman1999viabilityUniversalQC}. This implies that circuits composed entirely of Clifford gates can directly trade circuit depth for circuit width by using teleportation. Any corrective Pauli operations indicated by the measurement outcomes can be commuted in software through subsequent Clifford operations, and applied at the end of the circuit.\footnote{Note that while it is not necessary to wait the reaction time $\tau_r$ between teleporting in each Clifford gate, at the end of the circuit we must apply the overall required Pauli correction to the measurement outcome. Classically computing this correction may depend weakly on the number of Clifford gates applied. However, this additional run time is purely classical, and the calculation could likely be parallelized, and so the increase in total run time can likely be neglected.} We show this for teleporting in the $S$ gate using the magic state $\ket{S} = S\ket{+} = \frac{1}{\sqrt{2}}(\ket{0} + i \ket{1})$, where the conditional $Z$ correction can be commuted in software through the subsequent Hadamard gate:\\
\begin{center}
\begin{quantikz}
    \lstick{$\ket{\psi}$} & \ctrl{1} & \gate{H} & & \gate{X} & \\
    \lstick{$\ket{S}$} & \targ{} & \meter{} &\setwiretype{c}  & \wire[u][1]{c} \\[0.5cm]
\end{quantikz}
\end{center}

For circuits with non-Clifford gates, such as the $T$ gate, it is not possible to purely trade circuit depth for circuit width~\cite{fowler2012time,seilinger2013Tdepth}. Implementing a $T$ gate via magic state teleportation requires a Clifford correction conditioned upon the logical measurement result. As discussed in Sec.~\ref{Subsubsec:ReactionTime}, the time to infer the logical measurement and start the conditional operation is known as the reaction time, $\tau_r$. Using gate teleportation, we can implement $N$ sequential non-Clifford gates in time $N\tau_r$. This is known as reaction-limited, or time-optimal, quantum computation~\cite{fowler2012time}. The key building block is the auto-corrected gate, developed for both $T$~\cite{litinski2019gameofsurfacecodes} and CCZ~\cite{gidney2019flexible} gates, which teleport in the Clifford correction at the same time as the non-Clifford gate. The autocorrected $T$-gate is implemented by:
\begin{center}
\begin{quantikz}
    \lstick{$\ket{\psi}$} & \ctrl{2} & \gate{H} & & & & \gate{X} & \\
    \lstick{$\ket{T}$} & \targ{} & \meter{} &\setwiretype{c}  & \wire[d][1]{c} \\
    \lstick{$\ket{S}$} & \targ{} & & & \meter{X|Z} &\setwiretype{c} & \wire[u][2]{c}
\end{quantikz}
\end{center}

The measurement outcome of the $\ket{T}$ qubit determines whether the Clifford correction is needed. Application of the Clifford correction is controlled by the measurement basis of the $\ket{S}$ qubit. When the Clifford correction qubit is measured it `locks in' whether the correction is applied or not. The measurement outcome of $\ket{S}$ may indicate a residual Pauli $Z$ correction, which can be commuted through subsequent gates in software. In the diagram above, this means that there is no reaction time delay between applying the $T$ gate and applying the subsequent $H$ gate, in contrast to the standard circuit for $T$ gate teleportation. Below we show an example for the circuit $HTHT\ket{\psi}$~\cite{gidney2024Stack}:
\begin{center}
    \begin{quantikz}
    \lstick{$\ket{\psi}$} & & \ctrl{3} & \meter{} & \setwiretype{c} & \wire[d][6]{c} \\
    \lstick{$\ket{T}$} & & \targ{} & \meter{} & \setwiretype{c} \wire[d][1]{c}  \\
    \lstick{$\ket{S}$} & & \targ{} & & \meter{} & \setwiretype{c} \wire[d][4]{c} \\
    \lstick[2]{$\ket{B}$} & & \targ{} & \meter{} & \setwiretype{c} & \\
    & \gate{H} & \ctrl{3} & \meter{} & \setwiretype{c} & & & \wire[d][4]{c} \\
    \lstick{$\ket{T}$} & & \targ{} & \meter{} & \setwiretype{c} & \\
    \lstick{$\ket{S}$} & & \targ{} & & & \meter{} & \setwiretype{c} & \\
    \lstick[2]{$\ket{B}$} &  & \targ{} & \meter{} & \setwiretype{c} & & & \\
    & \gate{H} & &  & & & & \gate{P} &
\end{quantikz}
\end{center}
where $\ket{B} =\frac{1}{\sqrt{2}}(\ket{00}+\ket{11})$ denotes a Bell pair, and $P$ represents the final Pauli correction. Pauli corrections can potentially change future $T$ gates. For example a $Z$ correction is changed by the $H$ gate to an $X$ correction, which changes a subsequent $T$ to $T^\dag$. Such a change can be handled by updating our interpretation of the $\ket{T}$ qubit measurement result, and whether to apply an $S$ correction or not---controlled by the measurement basis of the $\ket{S}$ qubit. Because the teleported $S$ gate can lead to a $Z$ correction, the above argument shows that each preceding $\ket{S}$ logical measurement result must be known to determine the measurement basis of the subsequent $\ket{S}$ qubits. The measurements governing completion of each auto-corrected gate are thus separated by the reaction time $\tau_r$, which places a fundamental limit on the speed of this approach to FTQC.

We denote the time taken to prepare each gate teleportation module as $T_{\mathrm{Prep}}$. We saturate time-optimal computation when we have the resources to implement $G_{\mathrm{opt}}$ gates in parallel, where $G_{\mathrm{opt}}\tau_r = T_{\mathrm{Prep}}$. This ensures that the total reaction time delay associated with the gates is as much of a bottleneck as preparing the gate teleportation modules. While $T_{\mathrm{Prep}}$ may depend on the code distance $d$, we are able to implement $G_{\mathrm{opt}}$ logical operations in this time using autocorrected gate teleportation. The cost is amortized over the time to sequentially measure and decode the corrective qubits, resulting in an average time per non-Clifford gate of $\tau_r$. If the reaction time can be made faster than the time for lattice surgery (which requires $\mathcal{O}(d)$ rounds of syndrome extraction), this method can provide a large speedup over sequentially teleporting magic states via lattice surgery. If applied to all qubits,  reaction-limited computation approximately preserves spacetime volume~\cite{chamberland2022TwistFree} (although it may ultimately increase spacetime volume due to extra storage and routing requirements). However, if the circuit structure is such that many qubits are idle, then applying reaction-limited computation to the subset of active qubits could reduce the spacetime volume of the computation. The spacetime volume will also depend on the increased demand for magic states, and the requirement for sufficient routing/ancilla space to store the required Bell pairs and auto-corrected magic states. For example, the use of reaction-limited computation in Shor's algorithm in Ref.~\cite{gidney2021HowToFactor} was effective because of the pseudo-1D layout of the active addition registers, which enabled local magic state consumption and Bell pair creation.

\section{Resource estimate assumptions}
In our resource estimates presented in Sec.~\ref{Sec:FH_results}, we make the following assumptions:

\begin{itemize}
    \item Timescales: We assume a time of \SI{1}{\micro \second} for a round of syndrome extraction, and a reaction time $\tau_r =$ \SI{1}{\micro \second}. We assume a logical timestep of $d$ rounds of syndrome extraction, where $d$ is the code distance.
    \item Time for logical gates: The time for a CNOT, CZ, multitarget-CNOT, or multitarget-CZ gate is assumed to be two logical timesteps. The time for an $S$ gate is assumed to be one logical timestep. The time to teleport in a $T$-gate (equivalent to $Z(\pi/8)$ rotation up to global phase) is assumed to be one logical timestep plus one unit of reaction time\footnote{Using the compilation of~\cite[Fig.7]{litinski2019gameofsurfacecodes}, which is more efficient than teleportation executed with a lattice surgery CNOT~\cite{deBeaudrap2020ZXcalculusLatticeSurgery}.}. 
    The time to teleport in an auto-corrected $T$ or $\pi/8$ gate is assumed to be two logical timesteps plus one unit of reaction time. The time for a generic single-qubit Clifford gate is assumed to be two logical timesteps, on average.
    \item Routing: We include routing limitations imposed by conflicts in the target routing space between two parallel logical operations. However, we neglect routing limitations imposed solely by the orientation of patch boundaries (which would necessitate a rotation of the logical patch).
    \item Rotation synthesis: We assume that each rotation gate is synthesized to error $\epsilon_s$ using $\sigma =\lceil0.57\log_2(1/\epsilon_s) + 8.83 \rceil$ $T$ gates~\cite{Kliuchnikov2023shorterquantum}. The total synthesis error (it is assumed synthesis errors add linearly) is set as $1\%$ of the total algorithmic error (the remaining algorithmic error is allocated to error from the Hamiltonian simulation algorithm).
    \item Magic state distillation: Magic state factory (MSF) resources were taken from Ref.~\cite[Table 1]{litinski2019magicstate}. For $p=10^{-3}$ physical error, we adopt the $(15:1)^6_{13,5,5} \times (15:1)_{29,11,13}$ factory, which uses 39,100 physical qubits to output one magic state of fidelity $3.3 \times 10^{-14}$ in 97.5 rounds of syndrome extraction. For $p=10^{-4}$ physical error rates, we adopt the $(15:1)^4_{9,3,3} \times (20:4)_{15,7,9}$ factory, which uses 16,400 physical qubits to output 4 magic states of fidelity $2.4 \times 10^{-15}$ in 90 rounds of syndrome extraction. The total number of MSFs required is determined by specifying the rate of magic state consumption for the algorithm and increasing the number of MSFs until this rate is saturated. The target fidelity of magic state distillation is determined by assuming that the total error from imperfect $T$ gates (assuming errors add linearly) is less than 0.05 (i.e., we could have at most $1.5\times 10^{12}$ $T$ gates in the circuit at $p=10^{-3}$).
    \item The total logical spacetime volume $V$ is computed by multiplying the total number of logical qubits (i.e., not including qubits used for magic state distillation, but including routing space) by the circuit depth (total syndrome extraction depth + reaction time depth). $V$ is measured in logical-qubit $\cdot$ rounds of syndrome extraction. We assume that the logical error rate of each logical qubit per round of syndrome extraction, $p_L$, scales as~\cite{fowler2012SurfaceCodes,litinski2019gameofsurfacecodes}
    \begin{equation}
        p_L = 0.1\left(\frac{p}{p_*}\right)^{0.5(d+1)}
    \end{equation}
    where $p$ is the physical error rate, $p_*$ is the surface code threshold (assumed as 0.01 here), and $d$ is the code distance. We set the distance as the minimum value such that $V\cdot p_L \leq E$ for a specified failure rate $E$. The total number of physical qubits is given by the number of logical qubits multiplied by $2d^2$, plus the qubits for MSFs. The total wall time is given by the number of rounds of syndrome extraction plus the reaction depth.
\end{itemize}

\section{Resource estimates for plaquette Trotterization}\label{App:PlaqTrot}

\subsection{High-level implementation}\label{App:Subsec:HighLevel}

\begin{figure}
    \centering
    \includegraphics[width=0.8\linewidth]{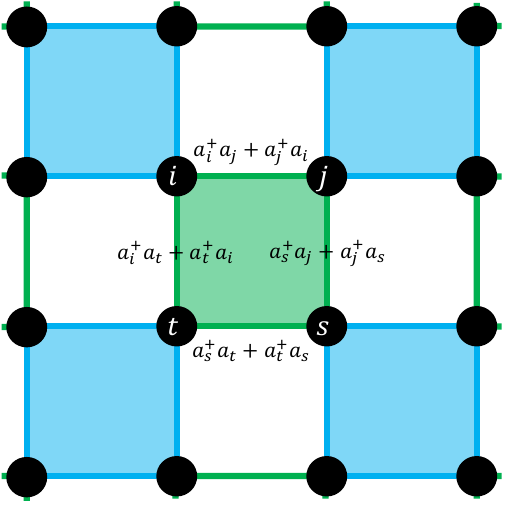}
    \caption{Coloring the Fermi-Hubbard lattice according to plaquette Trotterization (shown for a single spin). Hopping terms $a_x^\dagger a_{x'} + a_{x'}^\dagger a_x$ are grouped into mutually commuting `green' or `blue' plaquettes. The plaquette terms can be efficiently diagonalized, enabling simulation of time evolution under the hopping Hamiltonian in two groups. We highlight a single plaquette, indexed by sites $i,j,s,t$.}
    \label{Fig:PlaquetteDiagram}
\end{figure}

Ref.~\cite{CampbellHubbard22} introduced a new algorithmic compilation referred to as plaquette Trotterization (`PLAQ') for simulating the Fermi-Hubbard model. As shown in Fig.~\ref{Fig:PlaquetteDiagram}, the main idea is to divide the hopping terms in the Hamiltonian into two groups of `plaquettes', such that all the terms in each group commute. This division of the Hamiltonian enables tighter analytic bounds on the Trotter error (see App.~\ref{App:TrotterError}). Another beneficial feature of PLAQ is that it produces a circuit with many equal-angle rotations, allowing liberal use of Hamming weight phasing (HWP) to reduce the total number of $T$-gates required~\cite{gidney2018_halving_addition,Kivlichan2020ImprovedFaultTolerantSimulationCondensedMatter}. 

Time evolution is approximated using the 2nd-order Trotter formula $e^{iH\T} \approx \left(e^{iH_t^e \frac{\T}{2r}} e^{iH_t^o \frac{\T}{r}} e^{iH_t^e \frac{\T}{2r}} e^{iH_v \frac{\T}{r}} \right)^r$, where $H_v$ is the onsite interaction term, $H_t^{e/o}$ is the hopping Hamiltonian for even/odd plaquettes, and we have merged operations in the middle and start \& end of each 2nd-order Trotter step.

Time evolution under each plaquette operation, $e^{i \theta (T^\sigma_{ij} + T^\sigma_{js} + T^\sigma_{st} + T^\sigma_{ti} )} $ with $T^\sigma_{ij} = a_{i,\sigma}^\dag a_{j,\sigma} + a_{j,\sigma}^\dag a_{i,\sigma}$ (where $\theta$ depends on the total evolution time, number of Trotter steps, and hopping strength $t$), is implemented by diagonalizing the sum of terms around the plaquette. Each plaquette term is diagonalized using a gate sequence $V (\cdot) V^\dagger$, where $V= F_{jt}F_{is}F_{ij}$, and $F_{ij}$ is a two-mode fermionic Fourier transform (see definitions below). An important observation is that modes $i,s$ and $j,t$ are not adjacent in the fermionic mode ordering, and therefore under the Jordan-Wigner (JW) mapping these do not map to local gates acting on a constant number of fermionic modes. While alternative sequences of $F_{ij}$ operations can be used to diagonalize the plaquette operator, we were not able to find a sequence with gates that are local under the JW mapping. Locality can be restored using fermionic SWAP gates to reorder the modes~\cite{CampbellHubbard22}. It can be verified that the innermost $F_{ij}$ gate can omit the JW string of $Z$ operators, as the resulting phase error can be commuted through the diagonal rotation gates that follow, and cancels. Moreover, as discussed in Ref.~\cite[Appendix E]{CampbellHubbard22}, the combination of the innermost $F$ gate and rotations can be re-compiled to Clifford gates and rotations acting on disjoint qubits. In summary, the time evolution under each plaquette operation can be realized by $F_{jt} F_{is} C e^{i \theta Z_i} e^{i \theta Z_j} C^\dagger F_{is}^{\dagger} F_{jt}^{\dagger}$ where $C$ is a two-qubit Clifford gate.

In the JW mapping, we have $a_p = \frac{1}{2}\left(X_p + iY_p \right)\vec{Z}_{p,0}Z_0$, and $a_p^\dag = \frac{1}{2}\left(X_p - iY_p \right)\vec{Z}_{p,0}Z_0$, where $\vec{Z}_{pq}$ denotes Pauli $Z$ operators acting on all sites between $p$ and $q$ in the fermionic ordering. We compile the $F$ gates as follows. First observe that \cite[Eq I12]{babbush2018LowDepthQSimMaterial} provides a compilation of $F_{pq}$ equivalent to 
\begin{equation}
    F_{pq} = \exp\left(-\frac{i \pi n_p}{4}\right) \exp\left(\frac{i \pi n_q }{4}\right) \exp\left(\frac{i \pi f_s}{4}\right)
\end{equation} 
where $f_s = I + a_p^\dag a_q + a_q^\dag a_p - n_p - n_q$ is the fermionic SWAP gate (and $n_{p/q} = a^{\dag}_{p/q}a_{p/q}$ is the number operator for mode $p/q$), such that
\begin{align}
    F_{pq} a_p F_{pq}^\dag = \frac{1}{\sqrt{2}}\left(a_p + a_q\right) \\
    F_{pq} a_q F_{pq}^\dag = \frac{-i}{\sqrt{2}}\left(a_p - a_q\right).
\end{align}
Note that $F_{pq}[a_{p/q}^\dagger] = \left(F_{pq}[a_{p/q}]\right)^\dagger$. Second, observe that the desired relations in \cite[Eqs.~E6, E7]{CampbellHubbard22} still hold if the definitions in \cite[Eq.~E8]{CampbellHubbard22} are altered to match those above.
Next, under the JW mapping
\begin{equation}
    f_s \rightarrow \frac{1}{2}\left( X_p \vec{Z}_{pq} X_q + Y_p \vec{Z}_{pq} Y_q\right) + \frac{1}{2} \left(Z_p + Z_q \right).
\end{equation}
The two bracketed terms commute with each other, hence 
\begin{equation}
    \exp\left(\frac{i \pi f_s}{4}\right) = e^{\left(\frac{i \pi Z_p}{8}\right)} e^{\left(\frac{i \pi Z_q}{8}\right)} e^{\left(\frac{i \pi X_p \vec{Z}_{pq} X_q}{8}\right)} e^{\left(\frac{i \pi Y_p \vec{Z}_{pq} Y_q}{8}\right)}.
\end{equation}
Thus, up to an unphysical global phase,
\begin{equation}
    F_{pq} = e^{\left(\frac{i \pi Z_p}{4}\right)} e^{\left(\frac{i \pi X_p \vec{Z}_{pq} X_q}{8}\right)} e^{\left(\frac{i \pi Y_p \vec{Z}_{pq} Y_q}{8}\right)}.
\end{equation}
This is equivalent to two $\pi/8$ gates and one $S$ gate, and requires significantly fewer Clifford gates than the compilation in \cite[Fig.8]{Kivlichan2020ImprovedFaultTolerantSimulationCondensedMatter}.

As a result, implementing time evolution under a single plaquette operator requires 2 $R_z$ gates, 8 $T$-gates, 4 $S$-gates, and 2 two-qubit Clifford gates. As discussed in Ref.~\cite[Appendix E2]{CampbellHubbard22}, the plaquette operators share the same rotation angles, and so the number of $T$-gates required for synthesis can be reduced using HWP.

\subsection{Trotter error}\label{App:TrotterError}

We use the analysis in Ref.~\cite{CampbellHubbard22} to determine the number of Trotter steps required. As shown in \cite[Eq.~D6]{CampbellHubbard22}, the commutator bound that determines the Trotter error for plaquette Trotterization is given by:
\begin{equation}
    W_{\mathrm{PLAQ}} = W_{\mathrm{SO2}} + W_{\mathrm{Extra2}}
\end{equation}
where $W_{\mathrm{SO2}}$ is the commutator bound for the split-operator approach to Trotterization. Using \cite[Eq.~C4, Eq.~C32, Table III]{CampbellHubbard22} we find that 
\begin{equation}
    W_{\mathrm{SO2}} \leq \frac{U t^2 (2\sqrt{5}+16)L^2}{12} + \frac{1.5U^2 t L^2}{24}.
\end{equation}
Using \cite[Eq.~D10]{CampbellHubbard22}, $W_{\mathrm{Extra2}} \leq \frac{5}{12}L^2 t^3$.
Hence $W_{\mathrm{PLAQ}} \leq \kappa L^2 t^3$ with $\kappa = \frac{1}{24}\left(\frac{3}{2}\left(\frac{U}{t}\right)^2 + 2 \left(\frac{U}{t}\right)(2\sqrt{5}+16) + 10  \right)$. From Ref.~\cite{childs2021TheoryTrotter}, we use that $|| U(\T) - S_2(\T/r)^r || \leq \frac{W_{\mathrm{PLAQ}} \T^3 }{r^2} $, where $U(\T)$ is the time evolution operator for time $\T$, $S_2(\T/r)$ is the second-order Trotterization for time $\T/r$. Thus the number of Trotter steps required to achieve error $\epsilon$ is upper bounded by
\begin{equation}
    r \leq \frac{\kappa^{0.5} L (\T t)^{1.5}}{\epsilon^{0.5}}.
\end{equation}
For $\T = \mathcal{O}(L)$, we see that $r = \mathcal{O}(L^{2.5})$.

\subsection{Sequential $T$ gates}

The number of $T$ gates per Trotter step is determined in \cite[Appendix E]{CampbellHubbard22}. We quickly sketch the main ideas here. Following App.~\ref{App:Subsec:HighLevel}, we compute the $T$, Toffoli and rotation counts for each block in the Trotter step. 

The interaction term is straightforward. There are $L^2$ sites of each spin, and therefore $L^2$ onsite interactions to be applied. Each term of the form $e^{i \theta ZZ}$ can be implemented using two CNOT gates, and one arbitrary angle $Z$ rotation---and each term has the same angle. Using $\frac{L^2}{m}$ rounds of HWP with $m$ ancilla qubits, the $L^2$ rotations are compiled to approximately $L^2$ Toffoli gates, and $\frac{L^2}{m} \log_2(m)$ rotations. Each Toffoli gate is compiled to 4 $T$ gates, and each rotation is compiled to $\sigma$ $T$ gates. The total cost is thus $L^2\left(4 + \frac{\log_2(m) \sigma}{m} \right)$ $T$ gates.

The plaquette terms are costed similarly. Each individual plaquette requires two equal-angle rotation gates, and 4 $F$ gates. Each $F$ gate requires two $T$ gates. There are $\frac{L^2}{4}$ plaquettes of each color, per spin. Hence, for a given plaquette evolution in the Trotter ordering, we require $4L^2$ $T$ gates and $L^2$ rotations. Once again using $\frac{L^2}{m}$ rounds of HWP, and converting the Toffolis and rotations to $T$ gates, this yields $L^2\left(8 + \frac{\log_2(m) \sigma}{m} \right)$ $T$ gates.

Combining the interaction term and 3 plaquette terms, the total gate count is approximately $4L^2\left(7 + \frac{\log_2(m)\sigma}{m} \right)$ $T$ gates per Trotter step (see \cite[Eq.~E18]{CampbellHubbard22} for a slightly more precise estimate using more careful analysis of the costs of HWP). The total number of logical qubits is $2L^2 + m$.

We implement these $T$ gates using PBC, hence we only count non-Clifford gates~\cite{litinski2019gameofsurfacecodes}. PBC transforms $T$ gates to multiqubit $(\pi/8)$ rotations, which can be implemented by performing a joint lattice surgery operation between the qubits in the support of the $\pi/8$ rotation, and a $\ket{T}$ magic state. We assume that any corrective Clifford operations are performed in software, by updating the rotation axes of future $\pi/8$ gates (see~\cite{kim2022FaultTolerantQuantumChemicalSimulationsLiIon} for an in-depth discussion). 

Our choice of $m = \alpha L^2$ will mean that the HWP subroutine (the dominant cost in the algorithm) causes $\alpha L^2$ of the qubits to interact, leading to a rapid increase in the support of the $\pi/8$ gates. This justifies our assumption that $\pi/8$ gates are performed sequentially. This is in keeping with typical sequential circuits arising from PBC, and the original spirit of Ref.~\cite{CampbellHubbard22}.

We follow the PBC scheme of Ref.~\cite{litinski2019gameofsurfacecodes}, subsequently referred to as the `baseline architecture'~\cite{litinski2022activeVolume}. We adopt the `fast block' layout, which uses a large amount of routing space to ensure that one $\pi/8$ rotation can be applied each logical timestep. For $n$ logical qubits, this corresponds to $2n + \sqrt{8n} + 1$ surface code patches. We use sufficient MSFs to ensure that we can produce one $\ket{T}$ state per logical clock cycle.

We note that the scheme in Ref.~\cite{litinski2019gameofsurfacecodes} assumes that $Y$ basis lattice surgery measurements can be performed, which may require the use of more complex stabilizer measurement cycles. An alternative strategy is the approach of Ref.~\cite{chamberland2022TwistFree}, which avoids the use of twists, but introduces a factor of two in the time overhead of the computation. We ignore this subtlety in our resource estimates.

\subsection{$L$ gates in parallel}

\begin{figure*}
    \centering
    \includegraphics[width=0.9\linewidth]{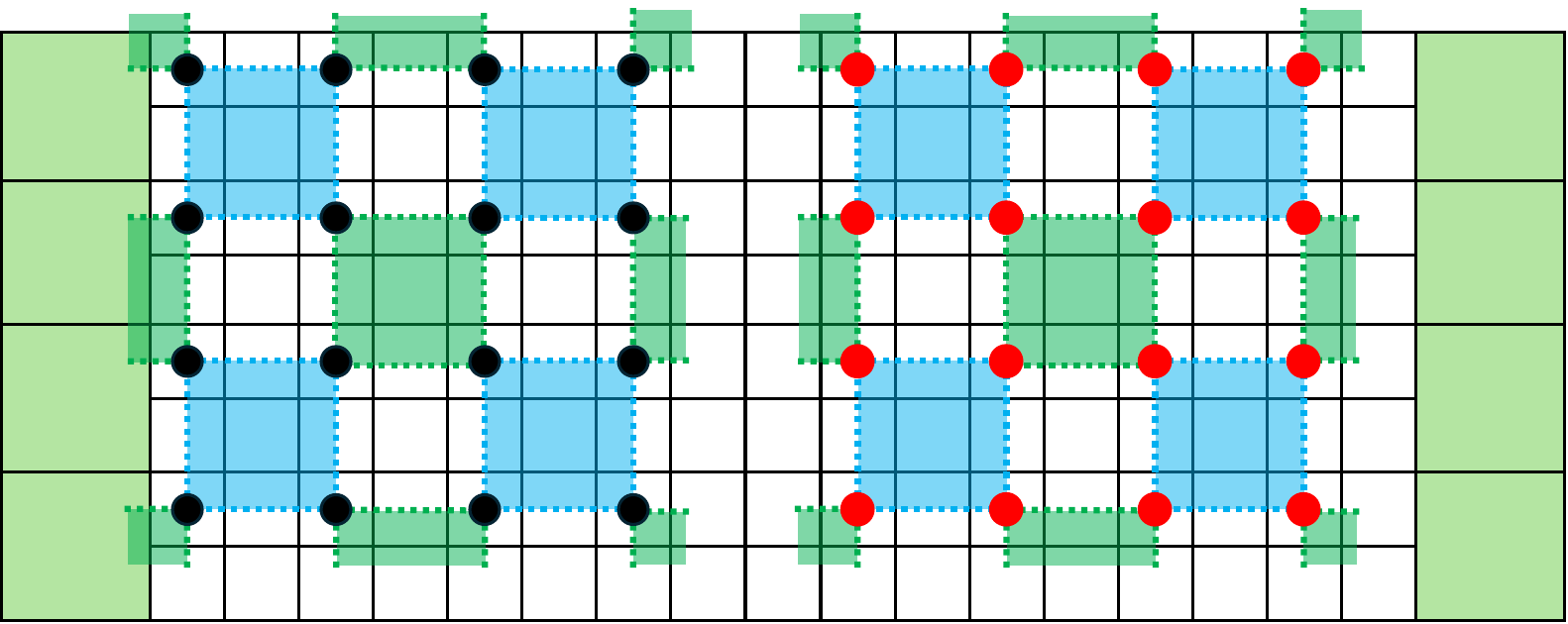}
    \caption{Surface code layout for the $L$-parallel compilation. White squares with black (red) circles denote spin up (down) fermionic sites. Empty white squares depict routing space. Light green squares along the boundaries depict magic state factories. The blue and green plaquettes are overlaid on the lattice, including periodic boundary conditions for each spin.}
    \label{Fig:L_Parallel_Layout}
\end{figure*}

The choice to apply $\mathcal{O}(L)$ $T$ gates in parallel is motivated by both the rate of magic state consumption from MSFs placed at the boundary of the compute area, as well as the non-locality of the fermionic operators under the Jordan-Wigner mapping. The calculation is laid out as shown in Fig.~\ref{Fig:L_Parallel_Layout}. We forgo HWP in order to save the depth required for computing the Hamming weight.  

We choose to implement Clifford gates explicitly, and assume that all $\pi/8$ gates require corrective $S$ operations, except when used for rotation synthesis (as the corrective Cliffords can be compiled to the end of the synthesis circuit).

There are $\frac{L^2}{4}$ plaquettes of a given color, per spin. Based on the layout in Fig.~\ref{Fig:L_Parallel_Layout}, the fermionic operators for a given plaquette operate over two fermionic rows under the Jordan-Wigner mapping. We implement $\frac{L}{2}$ non-overlapping plaquettes in parallel. Each plaquette requires 
\begin{itemize}
    \item Four $F$ gates (each requiring two multiqubit $\pi/8$ (and their corrective Clifford gates) and one $S$),
    \item Two CNOTs \& six single-qubit Clifford gates used to diagonalize the rotation gates such that they act on disjoint qubits (see \cite[Appendix E1]{CampbellHubbard22}),
    \item Two equal-angle $Z$ rotations, each synthesized from $\sigma$ $T$ states.
\end{itemize}
In rotation synthesis we only count the non-Clifford cost, motivated by compiling the Clifford gates in the synthesis circuit to the end of the synthesis circuit, as implemented in Ref.~\cite{wang2024optimizingFTQCtranspiler}. The rotations act on separate qubits, and so can be synthesized in parallel. As a result, we can implement $L/2$ plaquettes of each spin using
\begin{align}
    T~\text{Gate~Count:}&\quad \frac{L}{2} \times (8 + 2\sigma) \\
    \text{Depth:}&\quad 36 + \sigma
\end{align}
Hence, the total depth for implementing all terms in a plaquette operation is $L(18+\sigma/2)$. We require $L$ $\ket{T}$ states per spin sector per logical timestep.

Periodic boundary conditions provide an additional complication. Horizontal boundary interactions can be implemented in parallel using the strategy described above, as their JW strings do not overlap. However, the vertical boundary interactions would overlap in the JW mapping. We can implement the vertical boundary interactions by using $L$ layers of fermionic swap operations, each moving one periodic plaquette to a different row, such that they can then be implemented in parallel. The plaquettes are then swapped back to their original positions. The fermionic SWAP gate between sites $p$ and $q$ gate can be implemented by the circuit $\mathrm{CZ}_{pq} \mathrm{SWAP}_{pq} \mathrm{mCZ}_{q, \vec{pq}} \mathrm{mCZ}_{p, \vec{pq}}$. The operation $\mathrm{mCZ}_{q, \vec{pq}}$ denotes a multitarget $Z$ gate controlled on qubit $q$, targeting qubits between $p$ and $q$ in the JW ordering, and can be natively implemented in two logical timesteps using lattice surgery~\cite{fowler2018}. As a result, the vertical boundary interactions contribute an additional depth of $24L$ to the total depth.

The interaction term is more straightforward to implement. We implement the interactions between $2L$ sites and their opposite spin counterparts in parallel. We entangle the spins using two rounds of parallel CNOT gates across the partition. We synthesize $2L$ rotations in parallel, and then reverse the CNOT circuit. Overall, implementing the interaction term requires a depth of $L(4+\sigma/2)$. 

The total depth of a Trotter step is thus $L(2\sigma + 82)$ logical timesteps.

In each timestep, we consume at most $L$ magic states per spin. As a result, we require sufficient MSFs to produce $2L$ magic states per $d$ rounds of syndrome extraction. We consider a routing space overhead of $3:1$, in order to ensure clear paths between MSFs and data qubits, and for long-range fermionic swap operations.

\subsection{$L^2$ gates in parallel}

\begin{figure*}
    \centering
    \includegraphics[width=0.95\linewidth]{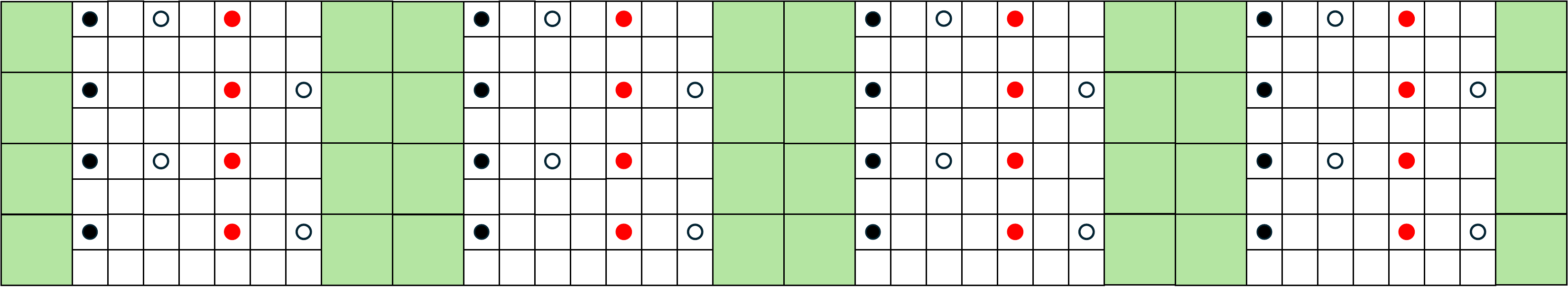}
    \caption{Surface code layout of the $L^2$-parallel compilation. White squares with black (red) circles denote spin up (down) fermionic sites. White circles denote the auxiliary qubits used in the local fermion-to-qubit mapping. Empty white squares depict routing space. Light green squares depict magic state factories. When factories are not distilling magic states, the space can be used for routing lattice surgery operations.}
    \label{Fig:L2_Parallel_layout}
\end{figure*}

\begin{figure*}
    \centering
    \includegraphics[scale=0.5]{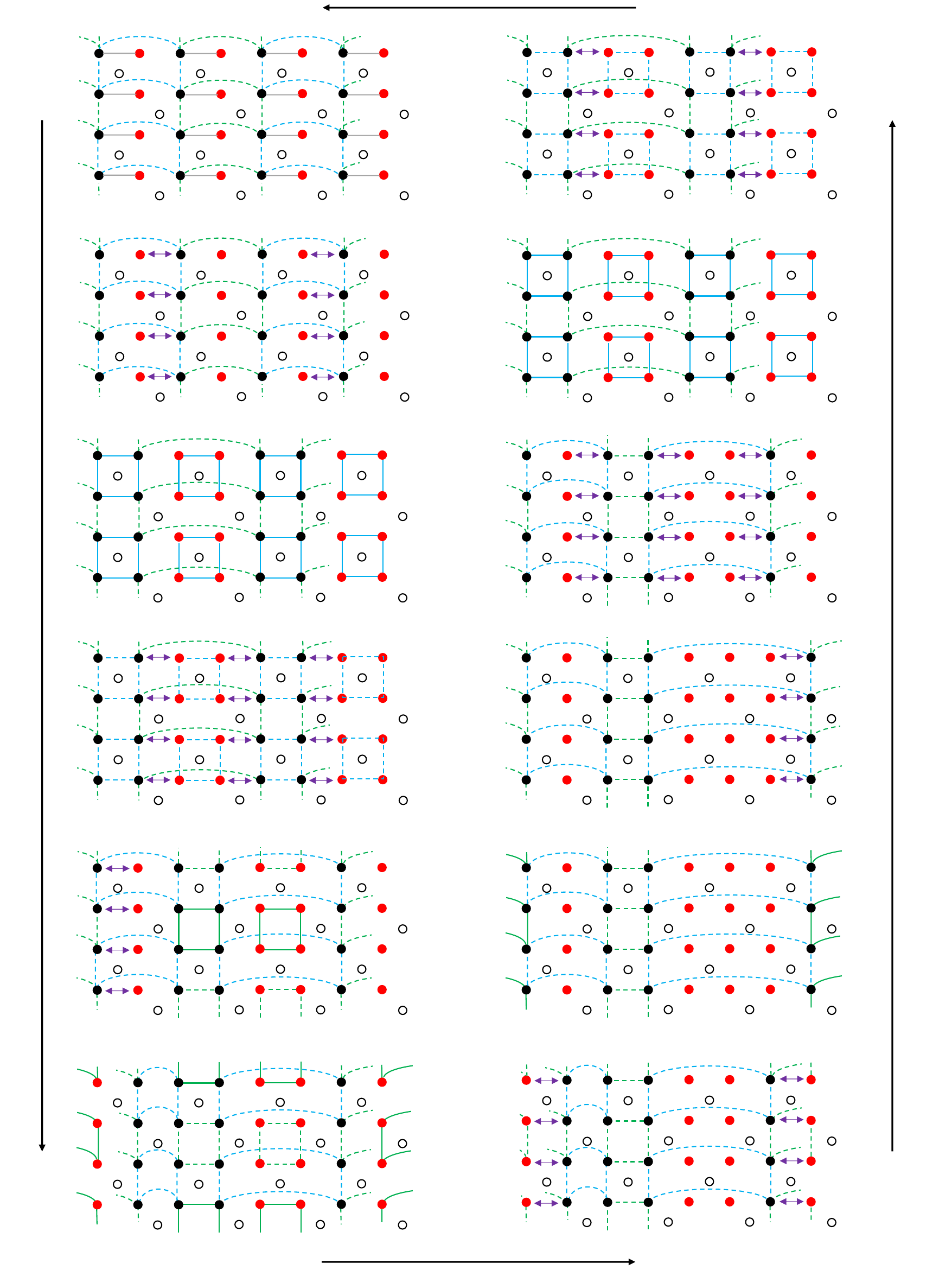}
    \caption{The sequence of operations for a bulk Trotter step using the compact mapping, starting in the top left. Spin up fermions are shown in black, and spin down in red. Open circles denote the ancilla qubits used in the compact mapping to provide locality. Double-headed purple arrows denote the use of fermionic swap gates between sites. Dashed blue and green lines indicate the plaquette interactions, and only a subset are included for clarity. Solid blue and green lines denote the implementation of the plaquette interactions, while solid grey lines denote implementation of the onsite interaction.}
    \label{fig:LocalFH}
\end{figure*}

In fermionic space, each of the plaquette operations are local, suggesting that $\mathcal{O}(L^2)$ (i.e. all plaquettes of a given color) can be implemented in parallel. In order to approach this level of parallelization, we must use a local fermion-to-qubit mapping, in order to eliminate the non-local JW strings that limit the available parallelism. In this work, we use the compact mapping, introduced in Ref.~\cite{derby2021CompactMapping}, and applied to NISQ simulation of the Fermi-Hubbard model in Ref.~\cite{clinton2021HamiltonianSimulationNearTermHubbard}. This mapping uses $1.5$ qubits per fermionic mode to ensure local fermionic operators are mapped to local qubit operators. The mapping is defined on an interaction graph, as shown in Ref.~\cite[Fig.~1]{derby2021CompactMapping}. The Fermi-Hubbard Hamiltonian is mapped to:
\begin{align}
    H &= H_{\mathrm{hop}}^h + H_{\mathrm{hop}}^v + H_{\mathrm{int}} \\
    H_{\mathrm{hop}}^h &= \frac{t}{2} \sum_{\langle i,j\rangle} (X_i X_j + Y_i Y_j) Y_{*(ij)} \\
    H_{\mathrm{hop}}^v &= \frac{t}{2} \sum_{\langle i,j\rangle } (-1)^{\delta_{ij,\uparrow}}(X_i X_j + Y_i Y_j) X_{*(ij)} \\
    H_{\mathrm{int}} &= U\sum_i Z_{i\uparrow}Z_{i\downarrow},
\end{align}
where $*(ij)$ denotes the auxiliary qubit adjacent to the oriented edge between sites $i,j$, and $\delta_{ij,\uparrow}$ is a Kronecker delta function denoting whether the oriented edge between sites $i,j$ is oriented upwards. We can lay out our Fermi-Hubbard lattice according to the interaction graph. We observe that the resulting plaquette operators are non-local in the interaction graph. In order to restore locality to the algorithm, we use local fermionic swap gates to re-order the fermionic modes during the algorithm. The sequence of operations for a Trotter step is shown in Fig.~\ref{fig:LocalFH}. 

Each fermionic swap gate along a vertical edge can be implemented by the sequence $\mathrm{CNOT}_{pq} \mathrm{CNOT}_{qp} \mathrm{CNOT}_{q*} \mathrm{CNOT}_{pq} \mathrm{CZ}_{pq}$. The horizontal fermionic swap is obtained by replacing $\mathrm{CNOT}_{q*}$ with $\mathrm{CY}_{q*}$. We assume that the fermionic swap gate can be implemented in 10 logical timesteps using lattice surgery. 

All gates within the plaquette operation are implemented sequentially, with the exception of the pair of rotations which are synthesized in parallel. The resulting depth is equivalent to 12 $\mathrm{CNOT}$ gates, $6$ single qubit Clifford gates, $8$ multiqubit~$\pi/8$ gates, $12$ $S$ gates, and $\sigma$ $T$ gates for rotation synthesis. 

The total depth of the Trotter step shown in Fig.~\ref{fig:LocalFH} is $6\sigma + 354$ logical timesteps. This depth assumes that we are able to consume $L^2$ magic states per logical timestep, which necessitates co-locating MSFs with data qubits. Each unit cell of 4 data qubits and one auxiliary qubit is served by two MSFs. As the MSFs may be much larger than a single surface code patch, it is necessary to distort the lattice to account for the MSFs. We initially explored leaving a buffer of routing space around each unit cell, but found that this contributed significantly to the footprint of the calculation. A more efficient approach for this calculation is to spin up the MSFs on demand. When not in use for distillation, the physical qubits in a factory can be used as routing space. Our circuit, shown in Fig.~\ref{fig:LocalFH} is separated into layers of Clifford horizontal fermionic swap gates, and interaction layers that act within a unit cell and contain non-Clifford gates. It is thus well suited to using the factory space for routing during the fermionic swap layers, and using the space for distillation during the interaction layers. We model the number of logical qubits that must be protected against faults as 
\begin{equation}
    4 \times 3 L^2 + f_r L^2 \left\lceil \frac{q_f}{2d^2} \left\lceil \frac{\tau_f}{\tau_m} \right\rceil\right\rceil,
\end{equation}
where $\tau_f$ is the time for the factory to distill a magic state, $\tau_m$ is the time between magic state consumption in the circuit, $q_f$ is the number of physical qubits in the factory, $d$ is the distance of the logical surface code patches, and $f_r$ is the fraction of time spent using the MSFs as routing space. The first term accounts for data qubits and the auxiliary qubits (and their routing space), and the second term accounts for the spacetime volume that the MSF regions spend as routing space (and so must be protected by the surface code). We make sure not to double count the factory qubits when reporting the final physical qubit counts. The layout is shown in Fig.~\ref{Fig:L2_Parallel_layout}.

\section{Resource estimates for quantum signal processing}\label{App:QSP}

\subsection{High-level implementation}

Quantum signal processing (QSP) approximates $e^{iH\T}$ by making repeated calls to a unitary block-encoding $U_H$ of the Hamiltonian $H$~\cite{low2016HamSimQSignProc,low2016HamSimQubitization}. The unitary $U_H$ is said to be an $(\alpha, m, \delta)$ block-encoding of $H$ if
\begin{equation}
    \left|\left| H - \alpha\left( \bra{0}^{\otimes m} \otimes I \right) U_H \left( \ket{0}^{\otimes m} \otimes I \right)  \right| \right| \leq \delta.
\end{equation}

QSP has optimal query complexity in terms of the evolution time $\T$ and the approximation error $\epsilon$, requiring~\cite[Corollary 60]{gilyen2018QSingValTransfArXiv} 
\begin{equation}
    \mathcal{O}\left(\alpha \T + \frac{\log\left(1/\epsilon\right)}{\log\left(e + \log\left(1/\epsilon\right)/\alpha \T\right)}  \right)
\end{equation}
calls to $U_H$. We use the bound in Ref.~\cite{babbush2019SYKmodel} which states that approximating $e^{iH\T}$ to error $\epsilon$ with QSP requires
\begin{equation}
    2\left(\alpha \T + \frac{3^{2/3}}{2}\left(\alpha \T\right)^{1/3} \log^{2/3}\left(1/\epsilon\right) \right)
\end{equation}
calls to $U_H$ (in our resource estimate we will neglect the small additional overhead resulting from multicontrol Toffoli gates required to implement the QSP phase factors, but include the cost of the necessary arbitrary-angle rotations~\cite{gilyen2018QSingValTransfArXiv}). We implement $U_H$ using the linear combination of unitaries (LCU) method, which decomposes $H = \sum_{j=0}^{L-1} c_j H_j$, with each $\norm{H_j} \leq 1$. We will work with the case where $H_j$ are unitary;\footnote{This holds, for example, if $H_j$ are multiqubit Pauli operators. If the $H_j$ are not unitary, a similar analysis holds given access to block-encoding unitaries $U_{H_j}$ for each $H_j$.} then, $H$ is an LCU, and a block-encoding unitary $U_H$ can be constructed such that 
\begin{align}
    (\bra{0}^{\otimes \ell} \otimes I) U_H (\ket{0}^{\otimes \ell} \otimes I) = \frac{H}{\alpha}\,,
\end{align}
where $\alpha =\sum_{j=0}^{L-1} |c_j|$, $\ell  = \lceil \log L \rceil$. The implementation of $U_H$ writes $U_H$ as a product of three unitaries
\begin{align}
    U_H = (\mathrm{PREPARE}^{\dag} \otimes I) \, \mathrm{SELECT} \, (\mathrm{PREPARE} \otimes I)\,.
\end{align}
The unitary PREPARE maps $\ket{0}^{\otimes \ell}$ to a superposition $\alpha^{-0.5}\sum_{j=0}^{L-1} \sqrt{|c_j|} \ket{j}$ encoding the Hamiltonian coefficients into its amplitudes. The unitary SELECT applies $H_j$ conditioned on the first $\ell$ qubits being in the state $\ket{j}$. That is,\footnote{To ensure SELECT is  unitary,   for all $j \in \{L,L+1,\ldots,2^{\ell} - 1\}$, we define $H_j$ to be the identity operator $I$ and $\mathrm{sign}(c_j)=1$. }
\begin{align}\label{eq:PREPSELPREP}
    \mathrm{SELECT} = \sum_{j=0}^{2^\ell-1} \ketbra{j} \otimes \mathrm{sign}(c_j)H_j\,.
\end{align}

The complexity of implementing PREPARE and SELECT for the Fermi-Hubbard model has been studied previously in the literature. Ref.~\cite{babbush2018EncodingElectronicSpectraLinearT} presented resource estimates for block-encoding the $N=2L^2$ site Fermi-Hubbard Hamiltonian, achieving $\alpha = (2t + 0.5U)N$, and requiring $\mathcal{O}\left( \log(N/\epsilon) \right)$ $T$ gates for PREPARE and $10N + \mathcal{O}\left(\log(N)\right)$ $T$ gates for SELECT. The circuit for SELECT dominates the cost, and has linear $T$-depth (the Clifford cost was not explicitly counted). In Ref.~\cite{yoshioka2022CondensedMatterSimulation} a tighter analysis of the same method achieved $\alpha = (2t + 0.125U)N$, and required $16\log(N)$ $T$ gates and 6 rotation gates for PREPARE. This latter work also investigated the use of parallelization to reduce the depth of the calculation, by using several copies of the PREPARE register to parallelize SELECT.

In our resource estimate we instead implement SELECT using the method of Ref.~\cite{Wan2021exponentiallyfaster}, which provided a $\log(N)$-depth implementation of SELECT, using no additional ancilla qubits. Our analysis requires some additional details not considered in Ref.~\cite{Wan2021exponentiallyfaster}, including the implementation of the crucial SWAPUP$^*$ subroutine in logarithmic depth for a restricted 2D connectivity, explicitly counting the prefactors of the Clifford gate depth, and converting between the PREPARE state of Ref.~\cite{babbush2018EncodingElectronicSpectraLinearT} and a PREPARE state suitable for the form of SELECT in Ref.~\cite{Wan2021exponentiallyfaster}. For $\T \in \mathcal{O}(L)$, the overall depth is $\tilde{\mathcal{O}}(L^3)$.

\subsection{Implementation of PREPARE}\label{Subsec:Prepare}
In Refs.~\cite{babbush2018EncodingElectronicSpectraLinearT,yoshioka2022CondensedMatterSimulation} the following state is used for $\mathrm{PREPARE}\ket{\bar{0}}$:
\begin{align}
    \sum_{p_x, p_y} &\sqrt{\frac{U}{4\alpha}} \ket{p_x, p_y \uparrow}\ket{p_x, p_y \downarrow}\ket{1}  \\ \nonumber
    + &\sqrt{\frac{t}{2\alpha}} \sum_{s \in \{\uparrow, \downarrow\}} \ket{p_x, p_y, s}\Big(\ket{p_x, p_y+1, s} + \ket{p_x, p_y-1, s} \\ \nonumber
    &+ \ket{p_x+1, p_y, s} + \ket{p_x-1, p_y, s}\Big)\ket{0},
\end{align}
which can be prepared with $16\log(N)$ $T$-gates and six arbitrary angle rotations. To be compatible with the SELECT oracle compilation of Ref.~\cite{Wan2021exponentiallyfaster} (see below), we work with a related PREPARE state, of the form
\begin{align}\label{Eq:OurPrep}
    \sum_i &\sqrt{\frac{U}{4\alpha}} \ket{i,\uparrow}\ket{i, \downarrow} \ket{00}\ket{0} \\ \nonumber
    + &\sqrt{\frac{t}{2\alpha}} \sum_{s \in \{\uparrow, \downarrow\}} \ket{i,s}(\ket{j,s} + \ket{k,s} \\ 
    + &\ket{\ell,s} + \ket{m,s})(\ket{10} + \ket{01})\ket{1},
\end{align}
where the sites $j,k,l,m$ are the nearest-neighbors in the lattice to site $i$. We can prepare this state by observing the following mapping between the two states:
\begin{align}
    &\ket{p_x, p_y, s} \nonumber\\
    \rightarrow& \ket{L p_y + (-1)^{p_y}p_x + (p_y \bmod 2)(L-1) + sL^2}
\end{align}
by associating $\uparrow=0, \downarrow=1$. This assumes a `snake ordering' of the lattice sites, and ordering all spin up sites before all spin down sites. As a result, we can first prepare the state from Refs.~\cite{babbush2018EncodingElectronicSpectraLinearT,yoshioka2022CondensedMatterSimulation} (step 1), and then convert it to the desired state (step 2) by:
\begin{itemize}
    \item Multiply $p_y$ by $L$ into a new register: Cost $\log^2(L)$ Toffolis.
    \item Controlled on the lowest bit of $p_y$, controlled-add/subtract~\cite{litinski2024Multiplication} $p_x$ into the new register: Cost $\log(L)$ Toffolis.
    \item Controlled on the lowest bit of $p_y$ add $L-1$ into the new register: Cost $2\log(L)$ Toffolis.
    \item Controlled on $s$, add $L^2$ into the new register: Cost $4\log(L)$ Toffolis. 
\end{itemize}
The total additional cost of step 2 is thus $\log^2(L) + 7\log(L)$ Toffolis, which is dominated by the cost of step 1. To simplify the analysis, we assume that PREPARE is implemented sequentially. The depth of sequential PREPARE is dominated by the depth of our SELECT oracle for all reasonable values of $L$, and so this contributes a negligible increase to our resource estimates.

\subsection{Implementation of SELECT}\label{Subsec:FH_SELECT}
The method of Ref.~\cite{Wan2021exponentiallyfaster} implements SELECT for a fermionic Hamiltonian $H = \sum_{pq} A_{pq} \left(a^\dagger_p a_q + a^\dagger_q a_p \right) $ mapped to qubits using the Jordan-Wigner mapping as
\begin{equation}
    H=\sum_{pq} \sum_{P_1 \in \{\pm X, \pm Y\}} \sum_{P_2 \in \{X, Y\}} c_{p,q,P_1, P_2} \left(P_1\right)_p \vec{Z}_{p,q} \left(P_2\right)_q.
\end{equation}
As noted in the original paper, the method is also applicable to more complex fermionic Hamiltonians, and other fermion-to-qubit mappings, such as the Bravyi-Kitaev mapping.
SELECT then acts as
\begin{equation}
    \mathrm{SELECT} \ket{p} \ket{q} \ket{P_1} \ket{P_2} \ket{\psi} = \ket{p} \ket{q} \ket{P_1} \ket{P_2} H_{pq}^{1,2}\ket{\psi},
\end{equation}
where $H_{pq}^{1,2} = \left(P_1\right)_p \vec{Z}_{p,q} \left(P_2\right)_q$. In our implementation we use $\ket{p}\ket{q}$ to index sites of the Fermi-Hubbard model (described in more detail in Sec.~\ref{Subsec:Prepare}). We use two qubits each for $\ket{P_1},\ket{P_2}$ with $\ket{10}\rightarrow X$ (implemented by a controlled $X$ gate) and $\ket{01} \rightarrow Y$ (implemented by a controlled $Y$ gate). This has the added benefit that by careful choice of the PREPARE state we can implement the interaction terms $-Z_p Z_q$ for no additional cost.

The high-level idea is to use a log-depth circuit SWAPUP$^*$, which controlled on a $\log(N)$ qubit register $\ket{x}$ swaps the $x$th qubit in $\ket{\psi}$ into the first position (and arbitrarily permutes the remaining qubits). By controlling on registers $\ket{p},\ket{q}$ the target qubits can be moved to the top of the register, and the required Pauli operations can be applied. The $Z$-string is implemented in a similar manner, by noting that the string results from operators $Z_p, Z_q$ conjugated by a log-depth ladder CNOT circuit (in our implementation we control $Z_p, Z_q$ on the final ancilla qubit in Eq.~\ref{Eq:OurPrep} so that the $Z$-string cancels for the interaction terms).

We note that the ladder CNOT circuit can still be implemented in log-depth on a processor with a 2D layout that realizes gates via lattice surgery, as the gates in a given layer do not overlap. This is not the case for the compilation of SWAPUP$^*$ presented in Ref.~\cite{Wan2021exponentiallyfaster}. In Fig.~\ref{fig:SWAPUP} we show an alternative compilation of SWAPUP$^*$, shown here for $n=8$ qubits, which still has log-depth for a 2D architecture with gates realized via lattice surgery.

\begin{figure}
    \centering
    \begin{quantikz}
    \lstick{A2} & & & & & & &  \ctrl{3} & \\
    \lstick{A1} & & & & & \ctrl{2} & \ctrl{6} & &  \\
    \lstick{A0} & \ctrl{1} & \ctrl{3} & \ctrl{5} & \ctrl{7} & & & & \\
    \lstick{Q0} & \swap{1} & & & & \swap{2} & & \swap{4} & \\
    \lstick{Q1} &  \targX{} & & & & & & & \\
    \lstick{Q2} & & \swap{1} & & & \targX{} & & &\\
    \lstick{Q3} & & \targX{} & & & & & & \\
    \lstick{Q4} & & & \swap{1} & & & \swap{2} & \targX{} & \\
    \lstick{Q5} & & & \targX{} & & & & & \\
    \lstick{Q6} & & & & \swap{1} & & \targX{} & & \\
    \lstick{Q7} & & & & \targX{} & & & & \\
\end{quantikz}
    \caption{A log-depth SWAPUP$^*$ network for $n=8$ qubits, constructed from controlled SWAP$^*$ gates (which have an imperfect phase that does not affect the block-encoding). $A$ denotes the control register qubits, and $Q$ denotes the target register qubits. The decomposition of the sequences of controlled SWAP$^*$ gates is shown in Fig.~\ref{fig:SWAPs}. The resulting circuit is compatible with a 2D layout where gates are implemented via lattice surgery (as gates do not overlap each other). The depth is $\lceil\log(N)\rceil$ sequences of controlled SWAP$^*$ gates.}
    \label{fig:SWAPUP}
\end{figure}

\begin{figure}
    \centering
    \begin{quantikz}
    \lstick{A0} & \ctrl{1} & \ctrl{3} & \ctrl{5} & \ctrl{7} & \\
    \lstick{Q0} & \swap{1} & & & & \\
    \lstick{Q1} &  \targX{} & & & & \\
    \lstick{Q2} & & \swap{1} & & & \\
    \lstick{Q3} & & \targX{} & & & \\
    \lstick{Q4} & & & \swap{1} & & \\
    \lstick{Q5} & & & \targX{} & & \\
    \lstick{Q6} & & & & \swap{1} & \\
    \lstick{Q7} & & & & \targX{} & \\
\end{quantikz}
=\begin{quantikz}
    & & & & & \ctrl{8} & & & & & \\
    & \targ{} & & \ctrl{1} & & & & \ctrl{1} & & \targ{} & \\
    &  \ctrl{-1} & \gate{A} & \targ{} & \gate{A} & \targ{} & \gate{A^\dagger} & \targ{} & \gate{A^\dagger} & \ctrl{-1} &  \\
    & \targ{} & & \ctrl{1} & & & & \ctrl{1} & & \targ{} & \\
    &  \ctrl{-1} & \gate{A} & \targ{} & \gate{A} & \targ{} & \gate{A^\dagger} & \targ{} & \gate{A^\dagger} & \ctrl{-1} &  \\
    & \targ{} & & \ctrl{1} & & & & \ctrl{1} & & \targ{} & \\
    &  \ctrl{-1} & \gate{A} & \targ{} & \gate{A} & \targ{} & \gate{A^\dagger} & \targ{} & \gate{A^\dagger} & \ctrl{-1} &  \\
    & \targ{} & & \ctrl{1} & & & & \ctrl{1} & & \targ{} & \\
    &  \ctrl{-1} & \gate{A} & \targ{} & \gate{A} & \targ{} & \gate{A^\dagger} & \targ{} & \gate{A^\dagger} & \ctrl{-1} &  \\
\end{quantikz}
    \caption{Decomposition of the sequence of controlled SWAP$^*$ gates, from Ref.~\cite{Wan2021exponentiallyfaster} (note that the SWAP$^*$ gates have an imperfect phase, which does not affect the block-encoding). Here $A = \exp\left(\frac{i\pi}{8} Y \right)$, equivalent to a $\pi/8$-gate, and we assume access to the multitarget CNOT gate through lattice surgery~\cite{fowler2018}. The depth is $5[\mathrm{CNOT}] + 4[\pi/8] + 4[S]$.}
    \label{fig:SWAPs}
\end{figure}
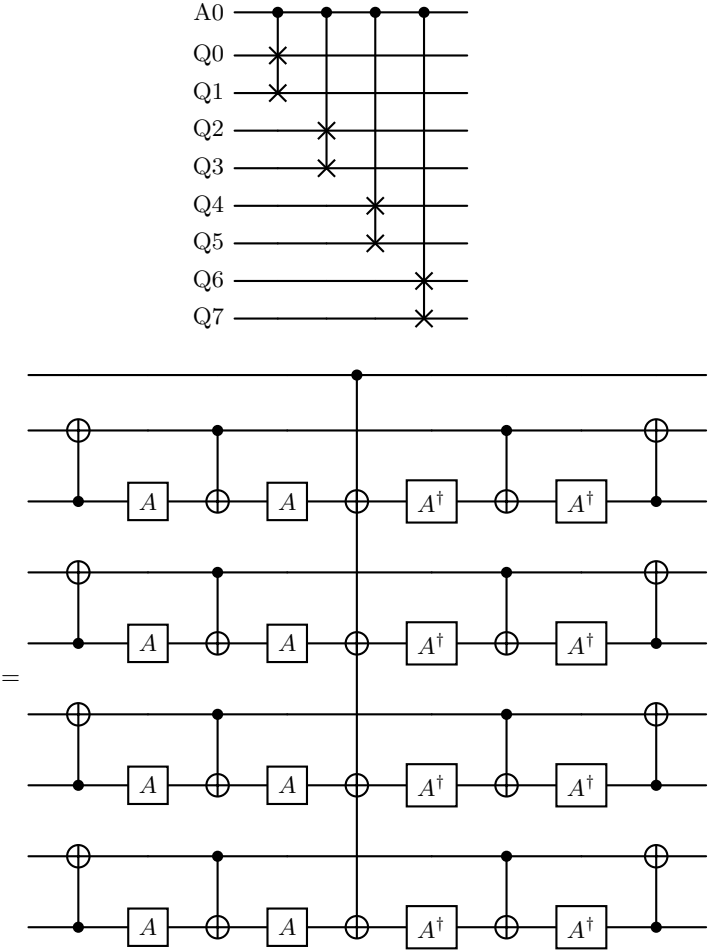

\begin{figure*}
    \centering
    \begin{quantikz}
        \lstick{$\ket{i}$} &  & \octrl{5}  & & \octrl{5} &  & & & & \\
        \lstick{$\ket{j}$} & &   & & & \octrl{4} & & \octrl{4} & & \\
        \lstick{$\ket{x}$} & &   & & &  & &  & &  \\
        \lstick{$\ket{y}$} & & &   & & &  & &  &  \\
        \lstick{$\ket{a}$} &  & & \ctrl{1} & & &  \ctrl{1} & & &  \\
        \lstick{$\ket{\psi}$} & \gate{L} & \gate{SU} & \gate{Z} & \gate{SU} & \gate{SU} & \gate{Z} & \gate{SU} & \gate{L} &  
    \end{quantikz}
    \caption{The first half of the SELECT circuit, which implements the Jordan-Wigner $Z$ string. For an input $\ket{i} \ket{j} \ket{x} \ket{y} \ket{a} \ket{\psi}$ the circuit implements $\ket{i} \ket{j} \ket{x} \ket{y} \ket{a} \vec{Z}_{ij}^a \ket{\psi}$, i.e. the $Z$ string between $i$ and $j$, controlled on the qubit $a$. In the diagram, open controls represent control on a register value (e.g. $\ket{i}$) rather than a binary value. The gate $L$ is the ladder of CNOTs, and $SU$ represents SWAPUP$^*$.}
    \label{fig:SEL_ZString}
\end{figure*}
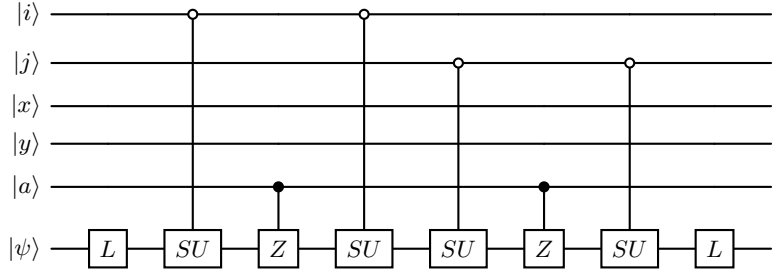

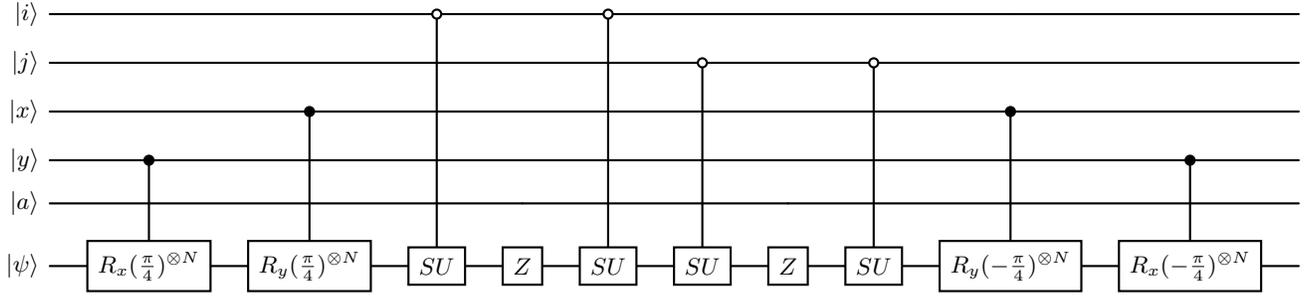
\begin{figure*}
    \centering
    \begin{quantikz}
        \lstick{$\ket{i}$} &  & & \octrl{5}  & & \octrl{5} &  & & & & & \\
        \lstick{$\ket{j}$} & &  &  & & & \octrl{4} & & \octrl{4} & & & \\
        \lstick{$\ket{x}$} & & \ctrl{3}  & & &  & &  & & \ctrl{3} & & \\
        \lstick{$\ket{y}$} & \ctrl{2} & &   & & &  & & & & \ctrl{2} & \\
        \lstick{$\ket{a}$} &  & &  & & &   & & & & &   \\
        \lstick{$\ket{\psi}$} & \gate{R_x(\frac{\pi}{4})^{\otimes N}} & \gate{R_y(\frac{\pi}{4})^{\otimes N}} & \gate{SU} & \gate{Z} & \gate{SU} & \gate{SU} & \gate{Z} & \gate{SU} & \gate{R_y(-\frac{\pi}{4})^{\otimes N}} &  \gate{R_x(-\frac{\pi}{4})^{\otimes N}} & 
    \end{quantikz}
    \caption{The second half of the SELECT circuit, which implements the hopping and interaction terms. For an input $\ket{i} \ket{j} \ket{1} \ket{0} \ket{a} \ket{\psi}$ the circuit implements $\ket{i} \ket{j} \ket{1} \ket{0} \ket{a} X_i X_j \ket{\psi}$. For an input $\ket{i} \ket{j} \ket{0} \ket{1} \ket{a} \ket{\psi}$ the circuit implements $\ket{i} \ket{j} \ket{0} \ket{1} \ket{a} Y_i Y_j \ket{\psi}$. For an input $\ket{i} \ket{j} \ket{0} \ket{0} \ket{a} \ket{\psi}$ the circuit implements $\ket{i} \ket{j} \ket{0} \ket{0} \ket{a} Z_i Z_j \ket{\psi}$. The first two terms (when combined with the JW string in Fig.~\ref{fig:SEL_ZString}) give the hopping term, while the 3rd term gives the onsite interaction term. In the diagram, open controls represent control on a register value (e.g. $\ket{i}$) rather than a binary value. The $SU$ gate represents SWAPUP$^*$.}
    \label{fig:SEL_int}
\end{figure*}

Our compilation of the SELECT oracle is presented in Fig.~\ref{fig:SEL_ZString} and Fig.~\ref{fig:SEL_int}, with additional optimizations over Ref.~\cite{Wan2021exponentiallyfaster} which effectively halve the $T$ depth quoted in \cite[Footnote 7]{Wan2021exponentiallyfaster}. The cost of SELECT is dominated by 8 calls to SWAPUP$^*$ (the $T$ count, $T$ depth and gate depth of SWAPUP$^*$ are shown in Table~\ref{tab:FH_SELECT_Resources}), and 2 calls to the CNOT ladder circuit (each with depth $2\lceil\log(N)\rceil-1$ CNOTs as explained in Ref.~\cite{Wan2021exponentiallyfaster}). Our optimized circuit also makes 4 calls to controlled-$R_{x/y}(\frac{\pi}{4})^{\otimes N}$, each implemented using two multitarget CZ gates, and $2N$ $\pi/8$ gates.

\begin{table}[h]
    \centering
    \begin{tabular}{c|c|c}
    $T$-count & $T$-depth & Gate depth \\ \hline
    $4(N-1)$ & $4\lceil\log(N)\rceil$ & $\lceil\log(N)\rceil\left(5[\mathrm{CNOT}] + 4 [\mathrm{\pi/8}] + 4[S] \right) $ 
    \end{tabular}
    \caption{Costs of implementing the SWAPUP$^*$ subroutine, which can be inferred from Figs.~\ref{fig:SWAPUP} \& \ref{fig:SWAPs}.}
    \label{tab:FH_SELECT_Resources}
\end{table}

\subsection{Resource estimation choices}
In order to saturate the $T$ consumption rate of SELECT, we co-locate MSFs with logical data qubits. We deliberately throttle the rate of magic state consumption to reduce the MSF overhead. For example, we can increase the depth of the SWAPUP$^*$ network from $\lceil\log(N)\rceil\left(5\mathrm{CNOT} + 4T\right) $ to $(\lceil\log(N)\rceil + 1)\left(5\mathrm{CNOT} + 4T\right) $ by splitting the first SWAPUP$^*$ sequence into two. This reduces the maximum number of parallel magic states required from $N/2$ to $N/4$, a large saving in MSF space for a modest depth increase. We use one MSF block per 4 logical data qubits, where the MSF block is capable of producing one $T$ state every 3 time steps. This value was chosen based on the structure of Fig.~\ref{fig:SWAPs}, which alternates between $T$-layers and CNOT layers. We use a routing overhead of $3:1$ for data qubits.

\end{document}